\documentclass[10pt,twocolumn]{article}

\usepackage[utf8]{inputenc}
\usepackage[T1]{fontenc}
\usepackage[a4paper,
            bindingoffset=5mm,
            left=15mm,
            right=15mm,
            top=15mm,
            bottom=15mm,
            footskip=5mm]{geometry}
\usepackage{amssymb}
\usepackage{mathtools}
\usepackage{graphicx}
\usepackage{verbatim}
\usepackage{xcolor}
\usepackage{tabularray}
\usepackage{subcaption}
\usepackage[hidelinks]{hyperref}
\usepackage{float}
\usepackage{physics}
\usepackage{siunitx}
\usepackage{cite}
\usepackage{booktabs}
\UseTblrLibrary{booktabs}    
\DefTblrTemplate{caption}{default}{}    
\DefTblrTemplate{capcont}{default}{}
\DefTblrTemplate{conthead}{default}{}    
\DefTblrTemplate{contfoot-text}{default}{\footnotesize{Continued on next column}}

\newcommand{\keywords}[1]{ {\small \textbf{\textit{Keywords:}} #1}}

\title{Deep Learning Models for ADITYA-U MHD Equilibrium}
\author{Udaya Maurya$^{1,2,a}$, Suman Aich$^{1,2}$, Indranil Bandyopadhyay$^{1,2,3}$, Daniel Raju$^{1,2}$ \\ \small $^1$Institute for Plasma Research, Bhat, Gandhinagar, India, 382428 \\ \small $^2$Homi Bhabha National Institute, Anushaktinagar, Mumbai, India, 400094 \\ \small $^3$ ITER-India, Institute for Plasma Research, Bhat, Gandhinagar 382428, Gujarat, India \\ \small $^a$ Author to whom correspondence should be addressed: udaya.maurya@ipr.res.in}
\hypersetup{pdfstartview={XYZ null null 1.00},
                      pdftitle={Deep Learning Models for ADITYA-U MHD Equilibrium},
                      pdfauthor={Udaya Maurya}}

\begin{document}
\maketitle
\begin{abstract}
This work presents deep learning models to predict magnetohydrodynamic equilibrium parameters and profiles for the ADITYA-U tokamak. A synthetic free-boundary equilibrium dataset consisting of 100,760 cases was generated using the pyIPREQ Grad-Shafranov solver, with inputs derived from 766 ADITYA-U plasma discharges and constrained to experimentally relevant circular limiter plasmas near the flat-top phase. Several deep learning approaches were investigated for predicting scalar equilibrium quantities, one-dimensional safety factor profiles and two-dimensional poloidal flux profiles. These approaches included Dense neural networks, principal component analysis based reduced-order models, one-dimensional and two-dimensional convolutional neural networks, and physics-informed neural networks incorporating Grad-Shafranov residual constraints. In addition, an inverse model was developed to estimate poloidal field coil currents from desired plasma equilibrium conditions. The results demonstrate that key equilibrium parameters and profiles can be accurately estimated within the operational domain represented by the dataset. The developed models provide a computationally efficient alternative to conventional equilibrium estimation and can be useful for real-time plasma control, rapid equilibrium analysis, and experimental planning in ADITYA-U operations.
\end{abstract}

\noindent \keywords{MHD Equilibrium, Deep Learning, Real-time Control, ADITYA-U Tokamak}

\section{Introduction}
\label{sec:intro}
Tokamak plasmas attain magnetohydrodynamic (MHD) equilibrium by balancing the plasma thermal pressure and the total magnetic pressure. MHD equilibrium is fundamental to tokamak research because several important quantities such as plasma boundary, magnetic axis position, safety factor, current density profile, Shafranov shift, internal inductance, etc. are directly derived from the equilibrium solution. These quantities play a central role in plasma control, transport and confinement studies, stability analysis, and the interpretation of experimental diagnostics. Under the assumption of axisymmetry, MHD equilibrium can be mathematically described by the Grad-Shafranov equation, a non-linear elliptic partial differential equation derived by balancing thermal and magnetic pressures. Consider a cylindrical coordinate system ($R$, $\phi$, $Z$) where $R$ represents the radial distance from the tokamak's central axis, $\phi$ is the toroidal angle, and $Z$ denotes the vertical coordinate. In terms of poloidal magnetic flux $\psi$, thermal pressure $p$ and toroidal flux function $F$ (related with toroidal magnetic field $B_{\phi}$ as $F=RB_{\phi}$), the Grad-Shafranov equation can be written in SI units as:
\begin{equation}
\frac{\partial^2\psi}{\partial R^2}+\frac{\partial^2\psi}{\partial Z^2}-\frac{1}{R}\frac{\partial\psi}{\partial R}=-\mu_0R^2\frac{\partial p}{\partial\psi}-F\frac{\partial F}{\partial\psi}
\label{eq:gs}
\end{equation}
The right side of the equation \ref{eq:gs} relates with toroidal current density $J_\phi$ as $J_\phi=Rp'+FF'/\mu_0R$. The left side is often denoted by the Grad-Shafranov operator $\Delta^*\psi$. Consequently, equation \ref{eq:gs} can be compactly written as $\Delta^*\psi=-\mu_0RJ_\phi$.

Equilibrium reconstruction is conventionally performed using iterative free-boundary solvers like EFIT\cite{1985Lao}, which are computationally expensive for applications requiring rapid or real-time evaluation. Fast surrogate models based on machine learning techniques have demonstrated reasonably good prediction performance while offering significant efficiency. Such approaches typically utilize synthetic equilibrium datasets generated for a specific tokamak and train machine learning models to learn the mapping between experimentally relevant inputs and equilibrium outputs. Restricting the dataset to a particular device enables the models to capture machine-specific operational characteristics while maintaining reliable predictive performance within the corresponding operating space. Studies like these have been reported for several tokamaks like KSTAR \cite{2020Joung}, NSTX-U \cite{2022Wai}, EAST \cite{2022Wan,2023Wan,2023Lu}, DIII-D \cite{2023Wei,2024Madireddy}, JET \cite{2026Rutigliano}, WEST \cite{2026Wan} and EXL-50U \cite{2025Zheng}. These studies explore various techniques like principal component analysis (PCA), convolutional neural networks (CNN), autoencoder architectures, transformer architectures, long short-term memory (LSTM), physics-informed neural networks (PINN), etc., demonstrating the potential of machine learning for fast equilibrium estimation and reconstruction.

ADITYA-U \cite{2024Tanna} is a medium-sized circular tokamak that serves as an important experimental platform for studies related to plasma equilibrium, transport, MHD activity and plasma control. Fast equilibrium estimation can be particularly useful for ADITYA-U operations because they can assist in rapid analysis of experimental discharges, actuator planning and real-time control applications. Motivated by these requirements, the present work develops deep learning models for predicting MHD equilibrium parameters and profiles representative of realistic ADITYA-U operating conditions. For this, a synthetic free-boundary equilibrium dataset consistent with experimentally observed ADITYA-U operational space has been generated using the free-boundary MHD equilibrium solver pyIPREQ \cite{2025Maurya,2025MauryaGit}. The generated dataset is constrained to circular limiter plasmas close to the flat-top duration and incorporates experimentally motivated operating ranges with physics-informed filtering criteria. Using this dataset, several deep learning models have been developed to predict scalar equilibrium quantities (Section \ref{sec:0D_Models}), one-dimensional safety factor profiles (Section \ref{sec:q}) and two-dimensional poloidal flux profiles (Section \ref{sec:psi}). The investigated approaches include dense neural networks, cascaded architectures, PCA-based reduced-order models, CNN-based models and PINN-constrained formulations. This work also presents a comparison between results of PCA models and CNN models. Physics-informed constraints are incorporated through Grad-Shafranov residual losses, while selected models also exploit known equilibrium properties such as the monotonic behavior of safety factor profiles. To the authors' knowledge, this is the first comprehensive study on machine learning based MHD equilibrium modeling for ADITYA-U using a large-scale synthetic database derived from experimentally relevant operating scenarios. The developed framework provides a foundation for future applications involving fast equilibrium inference, experiment planning and physics-guided real-time control for ADITYA-U operations.

\section{ADITYA-U Synthetic MHD Equilibrium Dataset}
\label{sec:data}
This section discusses the methodology to generate the synthetic MHD equilibrium dataset employed for training and evaluating the deep learning models and provides a statistical visualization of key physical quantities in the dataset. The objective was to generate a dataset representative of experimentally observed ADITYA-U operating conditions while minimizing the occurrence of physically unrealistic or potentially unstable equilibrium solutions. The generated dataset is limited to the circular limiter plasmas close to the flat-top phases of the discharges. Numerically, this dataset is restricted to the solutions of the Grad-Shafranov equation with an adopted current density profile (equation \ref{eq:Jprofile}).

The ADITYA-U MHD equilibrium dataset was generated using the free boundary equilibrium solver pyIPREQ \cite{2025Maurya,2025MauryaGit}. The input operating space was derived from 766 ADITYA-U plasma discharges spanning the years 2021-2025, corresponding to shot numbers 34227-38999. These discharges were selected based on the requirement that the plasma current $I_p$ exceeded \SI{100}{kA} for at least \SI{80}{ms}. The selected shots span several years of ADITYA-U operation and encompass a broad range of experimentally realized operating conditions. For each discharge, data was extracted from the duration near the flat-top phase, defined as the continuous time interval for which $I_p$ remained within 80\% of its peak value. In pyIPREQ, the computational domain consisted of a rectangular $R$-$Z$ grid with a spatial resolution of \SI{0.01}{m}, covering the ranges $R \in (\SI{0.4}{m}, \SI{1.1}{m})$ and $Z \in (\SI{-0.35}{m}, \SI{0.35}{m})$. Using this framework, 100,760 equilibrium cases were generated. Although the dataset contains over $10^5$ cases, samples originating from the same discharge and neighboring time instances are expected to exhibit significant correlation owing to the relatively slow evolution of plasma conditions in the considered operational duration. When generating this equilibrium dataset, eddy currents were ignored because the ADITYA-U vacuum vessel is electrically discontinuous at two toroidal locations \SI{180}{\degree} apart, and the considered operational duration generally has weaker temporal variations in plasma and coil currents compared with transient discharge phases. The poloidal field coil configuration used for equilibrium generation is summarized in Table \ref{tbl:ADITYAUcoils}.

{\centering
\footnotesize
\begin{longtblr}{
    colspec  = {l r r r r r},
    rowhead  = 0,        
    rowfoot  = 0,        
  }
\toprule
Coil & $R\,(m)$ & $Z\,(m)$ & $\Delta R\,(m)$ & $\Delta Z\,(m)$ & turns \\
\midrule
TR-1 & \num{0.22625} & \num{0} & \num{0.1075} & \num{1.04} & \num{174} \\
TR-2T & \num{0.3953} & \num{0.843} & \num{0.2305} & \num{0.16} & \num{56} \\
TR-2B & \num{0.3953} & \num{-0.83952} & \num{0.2305} & \num{0.157} & \num{56} \\
TR-3T & \num{1.22319} & \num{0.72806} & \num{0.03483} & \num{0.06} & \num{-3} \\
TR-3B & \num{1.226} & \num{-0.727} & \num{0.03516} & \num{0.0595} & \num{-3} \\
TR-4T & \num{1.534} & \num{0.60209} & \num{0.06446} & \num{0.0382} & \num{4} \\
TR-4B & \num{1.53007} & \num{-0.601} & \num{0.0654} & \num{0.03867} & \num{4} \\
TR-5T & \num{1.627} & \num{1.148} & \num{0.034} & \num{0.0155} & \num{-1} \\
TR-5B & \num{1.627} & \num{-1.1475} & \num{0.035} & \num{0.016} & \num{-1} \\
BV-1T & \num{0.37981} & \num{1.0527} & \num{0.16405} & \num{0.12003} & \num{-60} \\
BV-1B & \num{0.382} & \num{-1.051} & \num{0.16937} & \num{0.11733} & \num{-60} \\
BV-2T & \num{1.64204} & \num{1.1885} & \num{0.18155} & \num{0.03863} & \num{-22} \\
BV-2B & \num{1.641} & \num{-1.189} & \num{0.1795} & \num{0.038} & \num{-22} \\
MDI-T & \num{0.4625} & \num{0.297} & \num{0.055} & \num{0.065} & \num{12} \\
MDI-B & \num{0.462} & \num{-0.297} & \num{0.055} & \num{0.065} & \num{12} \\
MDO-T & \num{1.06288} & \num{0.3375} & \num{0.028} & \num{0.01312} & \num{1} \\
MDO-B & \num{1.06288} & \num{-0.3375} & \num{0.02814} & \num{0.01326} & \num{1} \\
ADI-T & \num{0.47} & \num{0.43} & \num{0.055} & \num{0.02166} & \num{4} \\
ADI-B & \num{0.4704} & \num{-0.43} & \num{0.055} & \num{0.02166} & \num{4} \\
FFI-T & \num{0.475} & \num{0.392} & \num{0.02943} & \num{0.0152} & \num{-1} \\
FFI-B & \num{0.47} & \num{-0.392} & \num{0.02923} & \num{0.01496} & \num{1} \\
FFO-T & \num{1.0875} & \num{0.377} & \num{0.02814} & \num{0.01291} & \num{1} \\
FFO-B & \num{1.0875} & \num{-0.377} & \num{0.02834} & \num{0.01335} & \num{1} \\
\bottomrule
\end{longtblr}
\addtocounter{table}{-1}
\captionof{table}{Poloidal field coil configuration adopted in the generation of the synthetic ADITYA-U free-boundary MHD equilibrium dataset. BCC coils and TF coils bus bars were ignored in the present work.}
\label{tbl:ADITYAUcoils}
}

In this work, $I_{OT}$ denotes the current per turn in all TR coils listed in Table \ref{tbl:ADITYAUcoils}, $I_{VF}$ denotes the current per turn in all BV coils, and $I_{FF}$ represents the current per turn in the FFI, FFO, and MDO coils (which are in series). The currents in the remaining ADI and MDI coils are collectively referred to as the divertor coil currents. The toroidal field coil current $I_{TF}$, which determines $B_{\phi}$, is also included in the study. ADITYA-U is equipped with sixteen magnetic probes that measure poloidal magnetic field $B_\theta$ \cite{2021Aich}. These probes are distributed in a poloidal cross-section, approximately along a full circular arc of radius \SI{0.275}{\meter} centered at (\SI{0.75}{\meter}, \SI{0.0}{\meter}). When generating the dataset, $B_\theta$ values at the locations of these probes were also computed from resultant $\psi$ profile, to be used as inputs to the deep learning models.

\subsection{Data Generation Methodology}
\label{sec:dtgen}
The pyIPREQ code requires specification of current density profile, which was customarily chosen to be:
\begin{equation}
J_\phi\propto\left(\beta\frac{R}{R_0}+\left(1-\beta\right)\frac{R_0}{R}\right)\left(1-\left(1-\bar\psi\right)^\alpha\right)^\gamma
\label{eq:Jprofile}
\end{equation}
where $\bar\psi$ represents normalized poloidal flux, defined such that $\bar\psi=1$ at the magnetic axis and $\bar\psi=0$ at the last closed flux surface (LCFS). The profile parameters $\alpha$, $\beta$ and $\gamma$ are determined using the procedure described below in this section.

The parameter $\beta$ is estimated from $\beta_p$ values available for few ADITYA-U discharges through Diamagnetic Loop measurements \cite{2024Aich}. Details regarding the $\beta_p$ data are discussed in Section \ref{sec:betap}. To generate equilibria consistent with prescribed $\beta_p$ values,  an additional step is added after step 6 of the algorithm mentioned in section 2.1 of \cite{2025Maurya}, or equivalently in the outer loop before Step G of the algorithm mentioned in Figure 2 of \cite{1979Johnson}. In this step, $\beta_p$ is computed using its standard definition $2\mu_0\langle p\rangle/\langle B_{pol}^2\rangle$. For the adopted $J_\phi$ profile, $p$ varies approximately linearly with the parameter $\beta$, allowing the required value of $\beta$ to be estimated iteratively from the target $\beta_p$.

\begin{figure}[!htb]
\begin{center}
\includegraphics[width=\columnwidth]{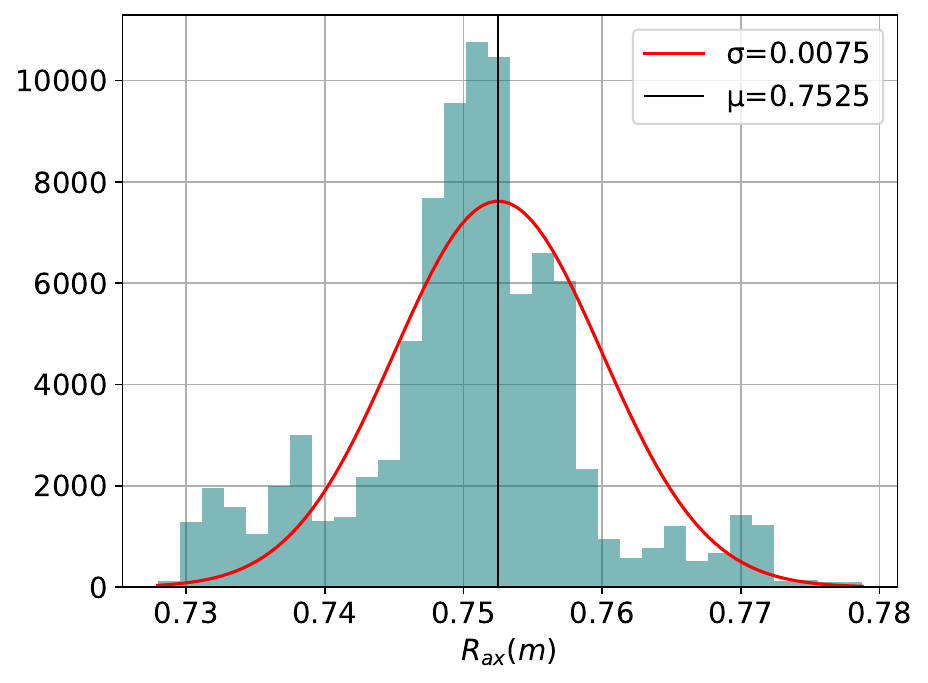}
\end{center}
\caption{Distribution of plasma current centroid position obtained from near flat-top phase of 16 ADITYA-U discharges. The overlaid normal distribution $\mathcal{N}(0.7525,0.0075)$ was adopted for sampling target magnetic-axis positions during equilibrium generation.}
\label{fig:Rax_dist}
\end{figure}

The profile parameters $\alpha$ and $\gamma$ may be determined using the plasma-position constrained optimization procedure described in section 3.1 of \cite{2025Maurya}. However, performing a full optimization procedure to estimate both parameters is computationally inefficient when generating a large equilibrium dataset. To improve efficiency, $\alpha$ is sampled directly and $\gamma$ is adjusted to match a target horizontal magnetic-axis position. The parameter $\gamma$ is estimated using Bayesian optimization with an Expected Improvement (EI) acquisition function by minimizing the objective function $\delta R_{ax}=|R_{ax,target}-R_{ax}|$. The optimization was limited to a maximum of ten equilibrium evaluations and terminated early whenever $\delta R_{ax}\leq\SI{0.01}{m}$. $R_{ax,target}$ was sampled from the distribution $R_{ax}\sim\mathcal{N}(0.7525,0.0075)$ (in meters), where $\mathcal{N}(\mu,\sigma)$ denotes the normal distribution with mean $\mu$ and standard deviation $\sigma$. This distribution was selected to approximately reproduce the range of plasma positions observed close to the flat-top phase of ADITYA-U discharges from Sine-Cosine diagnostic measurements \cite{2021Aich,2025Aich}, as shown in Figure \ref{fig:Rax_dist}. Although the Sine-Cosine diagnostic measures the plasma current centroid rather than the magnetic-axis position, the two quantities are generally close for circular low-$\beta$ plasmas. The parameter $\alpha$ was independently sampled from $\mathcal{N}(4.5,0.5)$, where $\mu$ and $\sigma$ values were empirically identified to produce equilibria with experimentally relevant core safety factor $q_0$ and convergence behavior.

In addition to the $J_\phi$ profile, pyIPREQ requires specification of the limiter boundary, poloidal field coil configuration, coil currents and $I_p$. $B_{\phi}$ is also required for calculating equilibrium quantities such as $q$ profile. The limiter boundary was prescribed as a circular contour of radius \SI{0.25}{\meter} centered at (\SI{0.75}{\meter},\SI{0.0}{\meter}). The coil currents for all the coils mentioned in Table \ref{tbl:ADITYAUcoils} were obtained from experimental measurements. Likewise, $I_p$ was available from measurements and $B_\phi$ was inferred from the measured $I_{TF}$. When equilibrium instances were generated, the input coil currents, $I_p$ and $B_\phi$ were smoothed in time and applied a 2\% uniformly random noise. Applying this noise helps in simulating potential measurement errors and broaden local operational sampling. Keeping the noise factor small ensures that the inputs are close enough for pyIPREQ to converge well and outputs are close to realistic scenarios.

Once the MHD equilibria were generated, specific filtering criteria were applied to determine whether to retain or reject each case. The maximum allowed deviation in the horizontal shift $\Delta R = R_{ax} - R_{maj}$ was \SI{0.04}{\meter}, and the vertical axis displacement $Z_{ax}$ was limited to \SI{0.02}{\meter}, aligning with experimental observations for ADITYA-U \cite{2021Aich,2025Aich}. The normalized internal inductance $\ell_i$ was constrained to the range $(0.7,1.7)$ in order to exclude equilibria with excessively broad or highly peaked current density profiles. The parameter $\gamma$ was additionally restricted to values below 8.9 to eliminate a small number of cases in which the Bayesian optimization procedure converged to unrealistically large values. The core safety factor $q_0$ was constrained to the range $(0.9,1.5)$, while the edge safety factor was required to satisfy $q_1>2.0$. These limits were selected to exclude extreme current-profile configurations and to retain equilibria representative of typical ADITYA-U operating conditions. Equilibria that failed to satisfy the Grad-Shafranov convergence criteria were also discarded. Although a formal linearized MHD stability analysis was not performed, the adopted filtering criteria help ensure that the final dataset consists of well-converged, physically plausible, and experimentally relevant equilibrium solutions.

\subsection{Probabilistic Estimation of Poloidal Beta}
\label{sec:betap}
\begin{figure}[!htb]
\begin{center}
\includegraphics[width=\columnwidth]{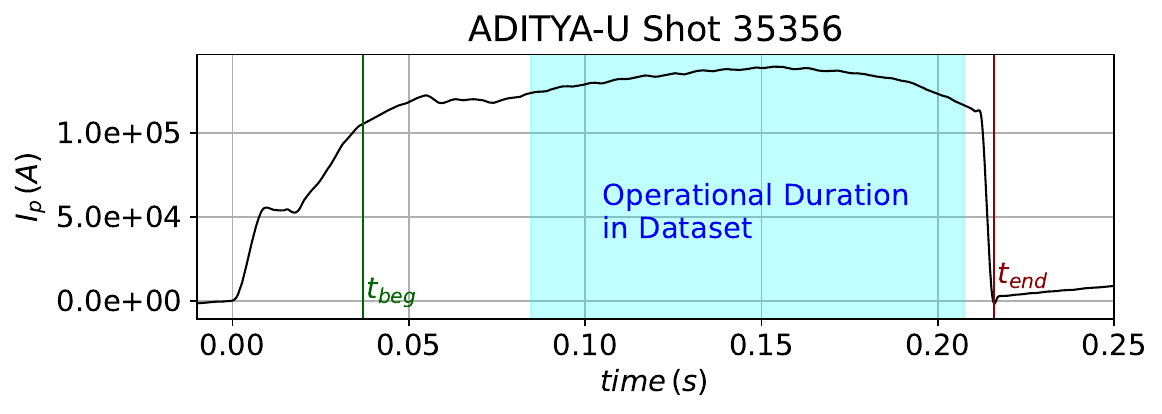}
\end{center}
\caption{$I_p$ and illustration of time stamps, for ADITYA-U shot 35356.}
\label{fig:Ip_time}
\end{figure}

As discussed in Section \ref{sec:dtgen}, $J_\phi$ profile parameter $\beta$ is determined from a target $\beta_p$ value. Since experimentally measured $\beta_p$ data are available only for few ADITYA-U discharges, a probabilistic surrogate model was developed to estimate $\beta_p$ for the all the 766 shots in the dataset. The objective of this model is not to provide an accurate diagnostic prediction of $\beta_p$, but rather to generate statistically plausible $\beta_p$ values for synthetic equilibrium generation when direct measurements are unavailable. Experimental $\beta_p$ measurements were available for 39 ADITYA-U discharges through Diamagnetic Loop diagnostics \cite{2024Aich}. To estimate $\beta_p$ for the complete dataset of 766 discharges, a linear regression model was developed using the inputs $I_p$, $I_{VF}$, $I_{OT}$, $I_{TF}$, $V_{loop}$ (loop voltage), and three time-dependent quantities. These quantities were the time at which the $I_p$ reached 75\% of its peak value ($t_{beg}$), the actual time instance under consideration ($t_i$), and the time at which $I_p$ returned to zero after the operational duration ($t_{end}$). Since $I_p$ starts rising at $t=0$ for each shot, the inverse of $t_{beg}$ signifies the rate at which $I_p$ rises. These quantities are illustrated in Figure \ref{fig:Ip_time} for ADITYA-U shot 35356. The motivation for using these time-based parameters was the absence of temperature and density measurements, which directly influence $\beta_p$.

\begin{figure}[!htb]
\begin{center}
\includegraphics[width=\columnwidth]{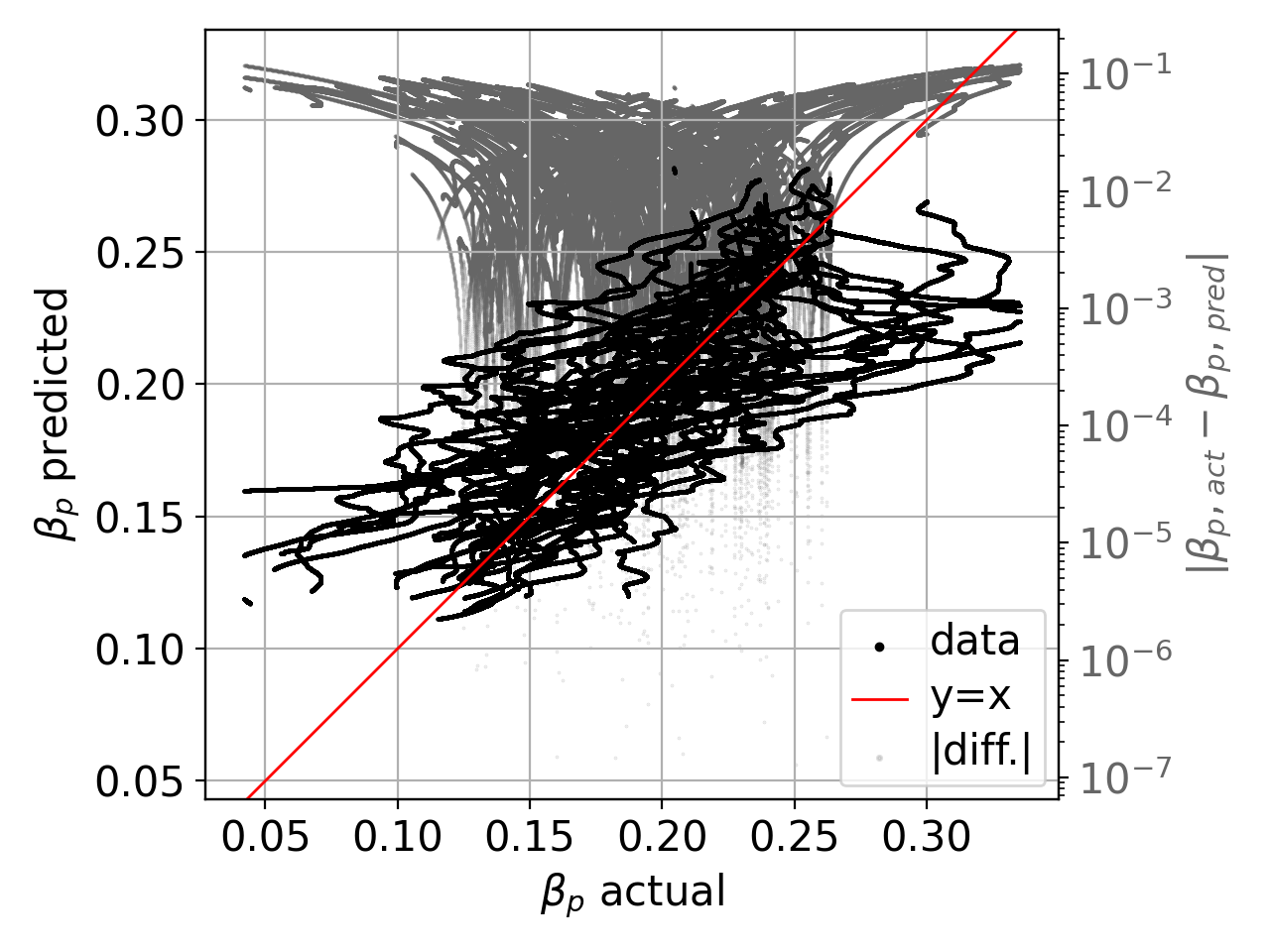}
\end{center}
\caption{Comparison between measured and predicted $\beta_p$ values on left axis, for the linear model discussed in Section \ref{sec:betap}. The right axis shows the absolute difference between them with median $\approx\num{0.0225}$ and $95^{th}$ percentile $\approx\num{0.0725}$.}
\label{fig:betap_lin_pred}
\end{figure}

Because the adopted model is linear, transformed variables that are expected to exhibit a more linear relationship with $\beta_p$ were used instead of the the raw quantities. The final inputs were $1/I_p^2$, $I_{VF}/I_p$, $I_{OT}/I_p$, $I_{TF}/I_p$, $V_{loop}/I_p$, $t_i/t_{end}$, $1/t_{beg}$, and $t_{end}$. Only instances near the flat-top phase were included during model development. Nonlinear dense neural-network models with 1-2 hidden layers were also investigated but they exhibited poorer extrapolation behavior on previously unseen discharges. Consequently, the linear model was selected because of its more stable and physically reasonable extrapolation characteristics. Figure \ref{fig:betap_lin_pred} compares the predicted and measured $\beta_p$ values for all 39 discharges, comprising 1,287,095 instances. This indicates inaccurate predictions and significant bias for high and low $\beta_p$ values. However, a highly accurate model is not necessary for the intended application. The model serves only as a stochastic prior used during equilibrium generation, where its role is to provide close to realistic estimates of the expected $\beta_p$ range and variability. Despite its limitations, the model provides a substantially more informed estimate than random sampling while remaining robust when applied to previously unseen discharges.

\begin{figure}[!htb]
\begin{center}
\includegraphics[width=\columnwidth]{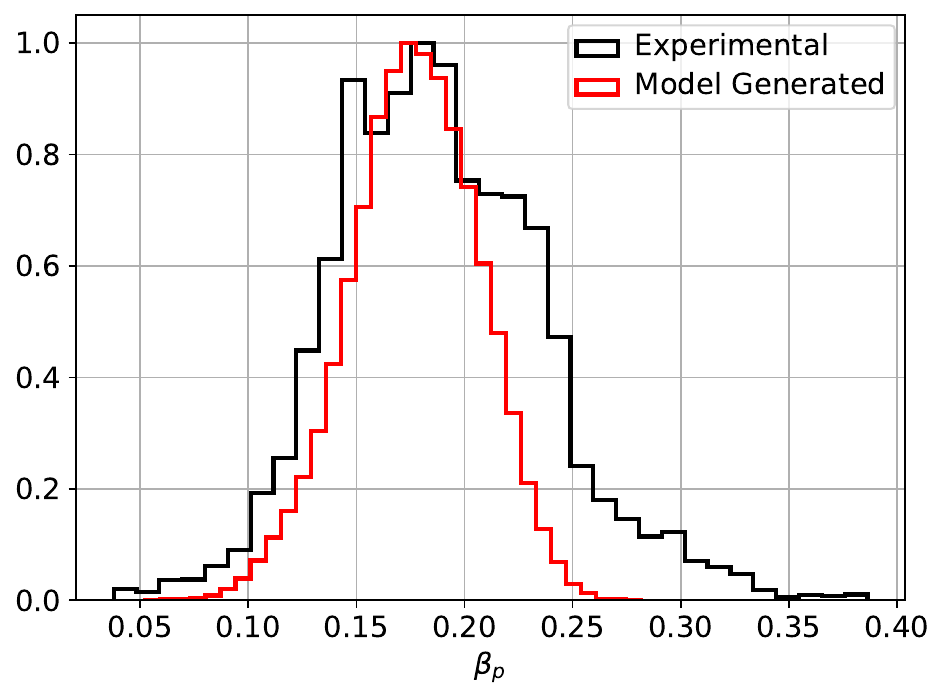}
\end{center}
\caption{Probability density distributions of experimentally measured $\beta_p$ values from 39 discharges and model-generated $\beta_p$ values for the 766-shot dataset. The density curves are normalized to maximum value 1 for visual comparison.}
\label{fig:betap_dist}
\end{figure}

To determine an appropriate noise to apply to this linear model, the residuals $(\beta_{p,act}-\beta_{p,pred})$ were analyzed and found to be reasonably approximated by a normal distribution $\mathcal{N}(-0.003,0.036)$. The robustness of this observation was evaluated using repeated shot-wise cross-validation. In each trial, 30 discharges were used for training and the remaining 9 discharges for testing. This procedure was repeated ten times, yielding residual distribution means within $(-0.026,0.007)$ and standard deviations within $(0.035,0.050)$. Based on these results, the final model was applied to the complete set of 766 discharges and the noise sampled from $\mathcal{N}(0.0,0.04)$ was applied during equilibrium generation in pyIPREQ. This stochastic perturbation preserves the uncertainty observed during cross-validation while broadening the sampled $\beta_p$ space available during equilibrium generation. The resulting $\beta_p$ values were clipped to the range $(0.05,0.4)$ based on experimentally observed operating conditions and to avoid generating nonphysical equilibria. Figure \ref{fig:betap_dist} compares the experimental $\beta_p$ distribution obtained from the 39 discharges with the distribution generated for the complete 766 discharges dataset. This further validates the use of this approach in estimating $\beta_p$ for equilibrium generation. The model-generated distribution is somewhat narrower than the experimentally observed distribution prior to noise injection. However, after incorporating the noise, the sampled $\beta_p$ values span a range comparable to that observed experimentally. It should be noted that the measured $\beta_p$ dataset also includes samples extending slightly beyond the selection criteria adopted for operational duration of the equilibrium dataset, which contributes to the broader experimental distribution.

\subsection{Data Visualization}
This subsection presents the statistical characteristics of the generated MHD equilibrium dataset. These distributions provide insight into the range of plasma conditions represented in the database and define the physical domain over which the deep learning models were trained and are expected to perform reliably. Summary statistics for selected equilibrium quantities are listed in Table \ref{tbl:vars_dist}. The distribution of $R_{ax}$ in the final dataset is slightly shifted inward relative to the target distribution shown in Figure~\ref{fig:Rax_dist}, which arises due to combined effects of equilibrium constraints, profile parameter selection, and subsequent filtering criteria applied during dataset generation. The $Z_{ax}$ is slightly negative for most cases, consistent with experimentally observed downward plasma shifts in ADITYA-U discharges \cite{2021Aich,2025Aich}.

\begin{figure}[!htb]
\begin{center}
\includegraphics[width=\columnwidth]{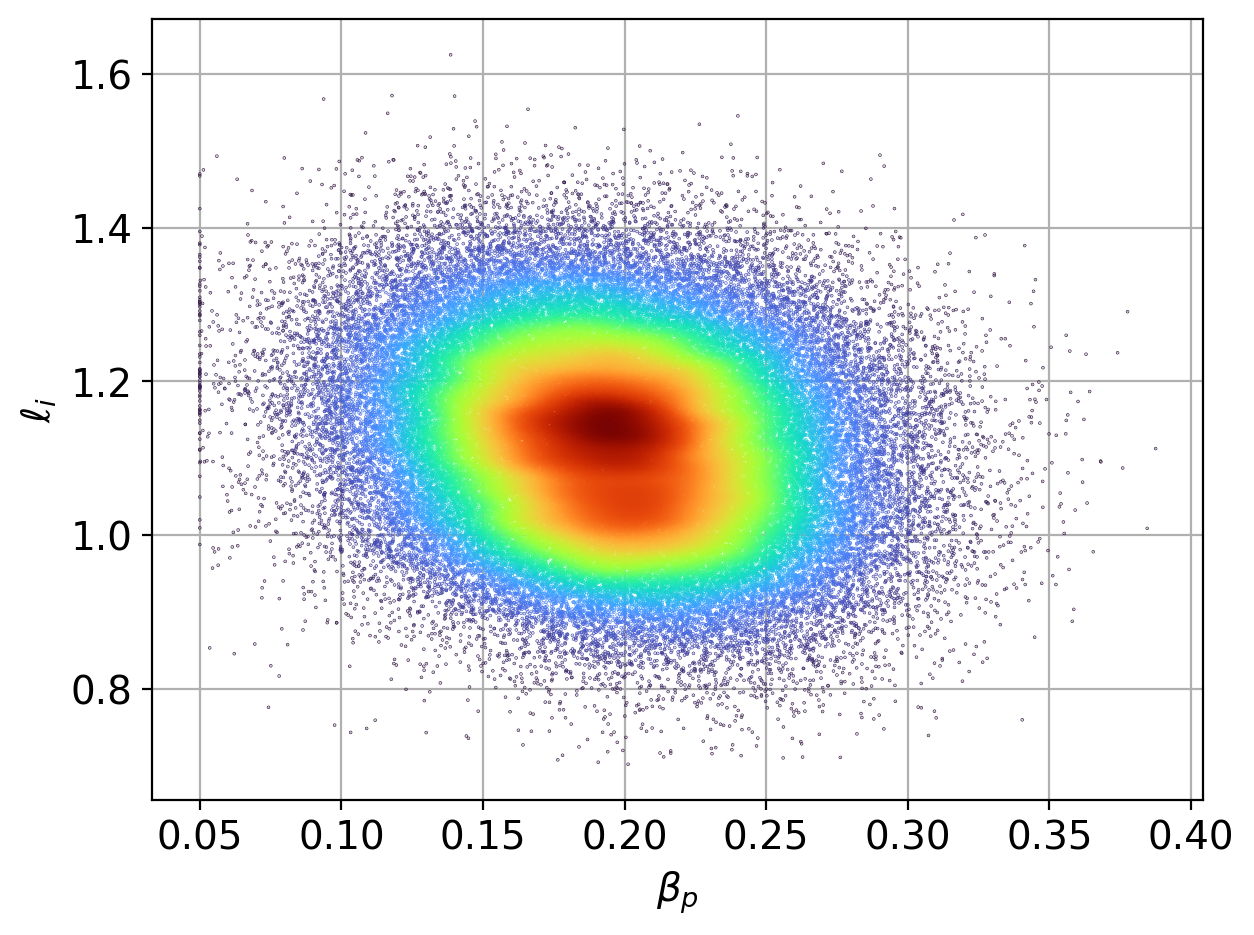}
\end{center}
\caption{Distribution of equilibria in the $(\beta_p,\ell_i)$ parameter space for all generated cases. Colors indicate the local density of data points.}
\label{fig:betap_li_dist}
\end{figure}

\begin{figure}[!htb]
\begin{center}
\includegraphics[width=\columnwidth]{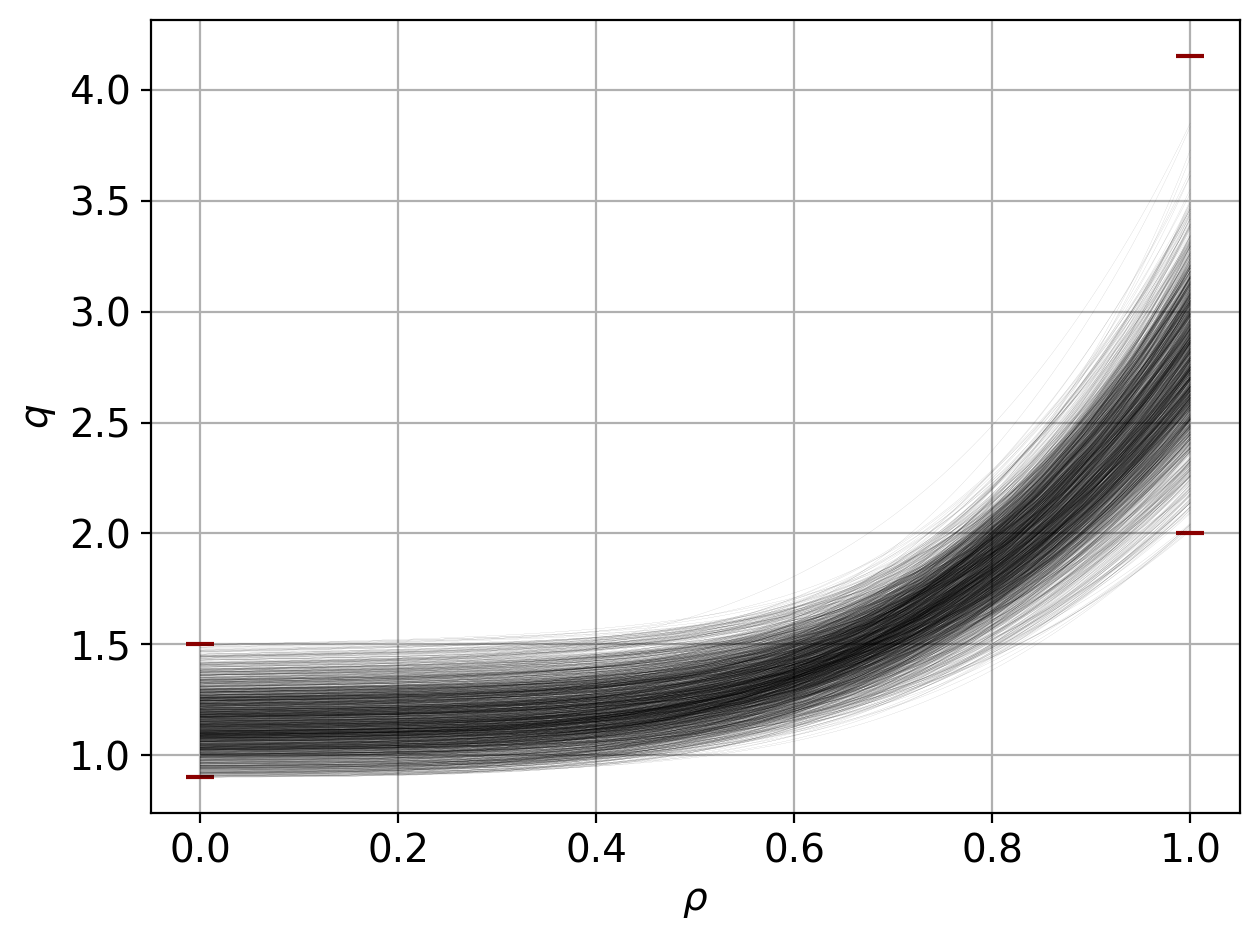}
\end{center}
\caption{Ensemble of $q$-profiles plotted with $\rho=1-\bar\psi$, for randomly selected 2000 cases from the dataset. The markers at $\rho=0$ and $\rho=1$ indicate ranges of $q_0$ and $q_1$ in the dataset.}
\label{fig:q_dist}
\end{figure}

\begin{figure}[!htb]
\begin{center}
\includegraphics[width=\columnwidth]{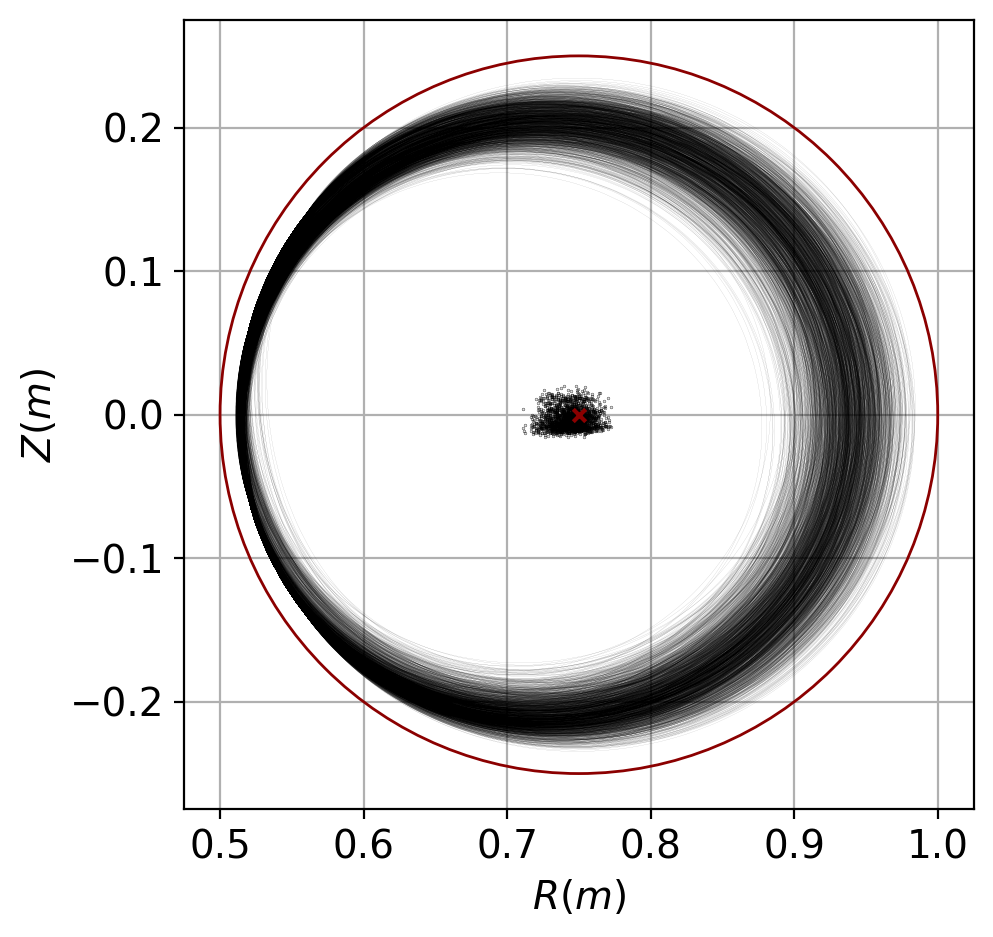}
\end{center}
\caption{Ensemble of LCFS and magnetic axis position for randomly selected 2000 cases from the dataset. The circular limiter boundary and machine's geometric center are also indicated for reference.}
\label{fig:lcfs_dist}
\end{figure}

{\centering
\small
\begin{longtblr}{
    colspec  = {l l l l},
    rowhead  = 0,        
    rowfoot  = 0,        
  }
\toprule
Variable & Median & 5-95\% & Min-Max \\
\midrule
$I_p\,(A)$ & \num{1.43e5} & \num{1.25e5}, \num{1.67e5} & \num{1.01e5}, \num{1.95e5} \\
$V_{loop}\,(V)$ & 2.189 & 1.509, 3.137 & -2.703, 6.174 \\
\footnotesize{$B_{\theta,MC}\,(T)$} & -0.104 & -0.130, -0.084 & -0.183, -0.062 \\
$P_{ax}\,(Pa)$ & 6.62e3 & 4.02e3, 9.94e3 & 1.19e3, 1.83e4 \\
$R_{ax}\,(m)$ & 0.744 & 0.726, 0.761 & 0.71, 0.778 \\
$Z_{ax}\,(m)$ & -0.00475 & -0.0116, 0.0111 & -0.0199, 0.02 \\
$\beta_p$ & 0.196 & 0.123, 0.27 & 0.05, 0.388 \\
$\ell_i$ & 1.118 & 0.929, 1.313 & 0.702, 1.625 \\
$\gamma$ & 5.626 & 3.323, 8.377 & 1.018, 8.9 \\
$q_0$ & 1.147 & 0.948, 1.385 & 0.9, 1.5 \\
$q_1$ & 2.779 & 2.304, 3.27 & 2.001, 4.153 \\
\bottomrule
\end{longtblr}
\addtocounter{table}{-1}
\captionof{table}{Statistical distribution of selected physical quantities in the equilibrium dataset. The column 5-95\% denotes the interval between the $5^{th}$ and $95^{th}$ percentiles. The quantity $B_{\theta,MC}$ represents $B_\theta$ evaluated at all 16 magnetic probe locations.}
\label{tbl:vars_dist}
}
Figure \ref{fig:betap_li_dist} shows the distribution of equilibria in the $(\beta_p,\ell_i)$ parameter space. This figure and Table \ref{tbl:vars_dist} show that $\beta_p$ in the dataset is in the range (0.05, 0.4) similar to the experimentally observed distribution shown in Figure \ref{fig:betap_dist}. Figure \ref{fig:q_dist} presents an ensemble of $q$-profiles for randomly selected 2000 cases in the dataset, illustrating the variability in current profiles in the database and the imposed constraints on $q_0$ and $q_1$. Figure \ref{fig:lcfs_dist} shows an ensemble of LCFS and magnetic axis positions for randomly selected 2000 cases in the dataset, illustrating the spread of plasma boundary and plasma position in the dataset. The small apparent gap between LCFS and the limiter boundary is due to the discretization of circular contour on finite computational grid with spacing \SI{0.01}{\meter}.

\section{Deep Learning Methodology}
\label{sec:DLmethod}
This section describes the common methodology adopted for training, hyperparameter optimization, and evaluation of the deep learning models presented in this work. All models were developed in Python-3 using the \texttt{tensorflow} framework with the \texttt{keras} API \cite{2015TensorFlow}. Hyperparameter optimization was performed using the \texttt{keras-tuner} library \cite{2019KerasTuner}. Equilibrium data generation, hyperparameter optimization and model training were carried out on a multi-core high-performing computing (HPC) facility using CPU-based parallel execution. Depending on model complexity, hyperparameter optimization typically required from several hours to multiple days. Four classes of machine learning approaches were investigated in this work:
\begin{enumerate}
\item Dense neural networks for modeling scalar equilibrium quantities and low-dimensional parameter mappings.
\item Principal Component Analysis (PCA) based models that reduce high dimensionality of $q$ and $\psi$ profiles to a lower-dimensional latent space.
\item Convolutional Neural Networks (CNN) for exploiting local spatial correlations and smoothness of $q$ and $\psi$ profiles.
\item Physics Informed Neural Networks (PINN)for incorporating Grad-Shafranov constraints into the prediction of $\psi$ profiles.
\end{enumerate}
For model development, the dataset was divided into training, validation, and testing subsets in the proportion 0.8:0.1:0.1. The training subset was used for parameter optimization, the validation subset was used for model selection, and the testing subset was reserved exclusively for final performance evaluation. The data split was performed based on shot labels rather than individual cases because multiple equilibria originating from the same discharge are temporally correlated and cannot be regarded as statistically independent samples. This strategy reduces the possibility of information leakage between training and testing datasets and provides a more realistic assessment of model generalization to previously unseen discharges. Statistical outlier filtering (like Z-score filtering) was not applied because the dataset already consisted of converged and physically plausible equilibrium cases, which allows the models to also learn any extreme cases present in the dataset. For dense neural-network models, all input and output variables were scaled to have mean $\mu=0$ and standard deviation $\sigma=1$, computed from the training dataset. The same transformation was subsequently applied to the validation and testing datasets to avoid any information leakage. For CNN-based models, the profile quantities were scaled as complete profiles rather than on a point-by-point basis. All models were trained using the \texttt{Adam} optimizer. Mean Squared Error (MSE) was used as the primary loss function and validation metric, except for the PINN models discussed in Section \ref{sec:psi}, which additionally included physics-based loss terms.

Hyperparameter optimization was performed using the \texttt{Hyperband} algorithm implemented in \texttt{keras-tuner}, together with variable learning rates and early stopping mechanism. The search space included hidden-layer activation functions \texttt{relu}, \texttt{elu}, \texttt{selu}, \texttt{tanh}, \texttt{silu}, \texttt{gelu} and \texttt{softplus}, which represent a mix of commonly used functions in physics based modeling and have variety of nonlinear and gradient characteristics. The search space for Dense models included number of hidden layers, layer widths, and dropout rates (where applicable). The hyperparameter search spaces for CNN-based models varied according to the specific application and are described in the corresponding sections. Following hyperparameter optimization, the selected architecture was retrained using larger epoch limits to obtain the final model. To ensure the performance consistency of the final model, it was trained and evaluated 2-3 times more on independently randomized shot-wise split data.

\section{Scalar Parameters Models}
\label{sec:0D_Models}
This section presents dense neural-network models developed for predicting key scalar MHD equilibrium quantities like magnetic axis position ($R_{ax},Z_{ax}$), poloidal beta ($\beta_p$), normalized internal inductance ($\ell_i$), edge safety factor ($q_1$), and the poloidal flux values at the magnetic axis and limiter ($\psi_{axs},\psi_{lim}$). In addition, an inverse model is developed to estimate poloidal field coil currents for desired equilibrium conditions. Besides their standalone utility, several of these models serve as intermediate components for the profile-prediction models discussed in Sections \ref{sec:q} and \ref{sec:psi}. Joint prediction of multiple equilibrium quantities was also investigated, including the combinations ($R_{ax},Z_{ax}$, $\beta_p$, $\ell_i$, $q_1$) and ($\beta_p$, $\ell_i$, $q_1$). However, these multi-output models generally exhibited reduced prediction accuracy compared with dedicated models trained for smaller groups mentioned in this section.

\subsection{Magnetic Axis Position Model}
\label{sec:RZax}
\begin{figure}[!htb]
\begin{center}
\includegraphics[width=\columnwidth]{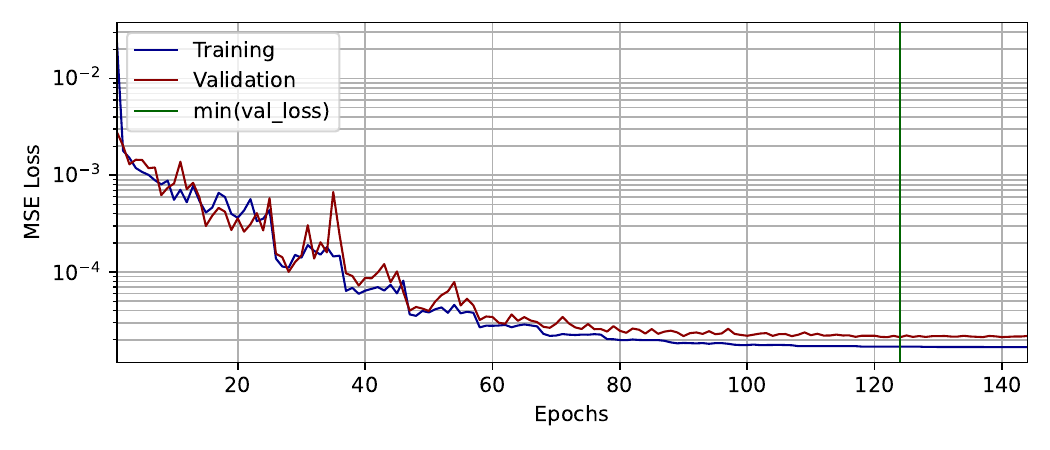}
\end{center}
\caption{MSE Loss for Training and Validation Datasets for model shown in Table \ref{tbl:RZax_model}.}
\label{fig:RZax_loss}
\end{figure}

\begin{figure}[!htb]
\begin{center}
\includegraphics[width=\columnwidth]{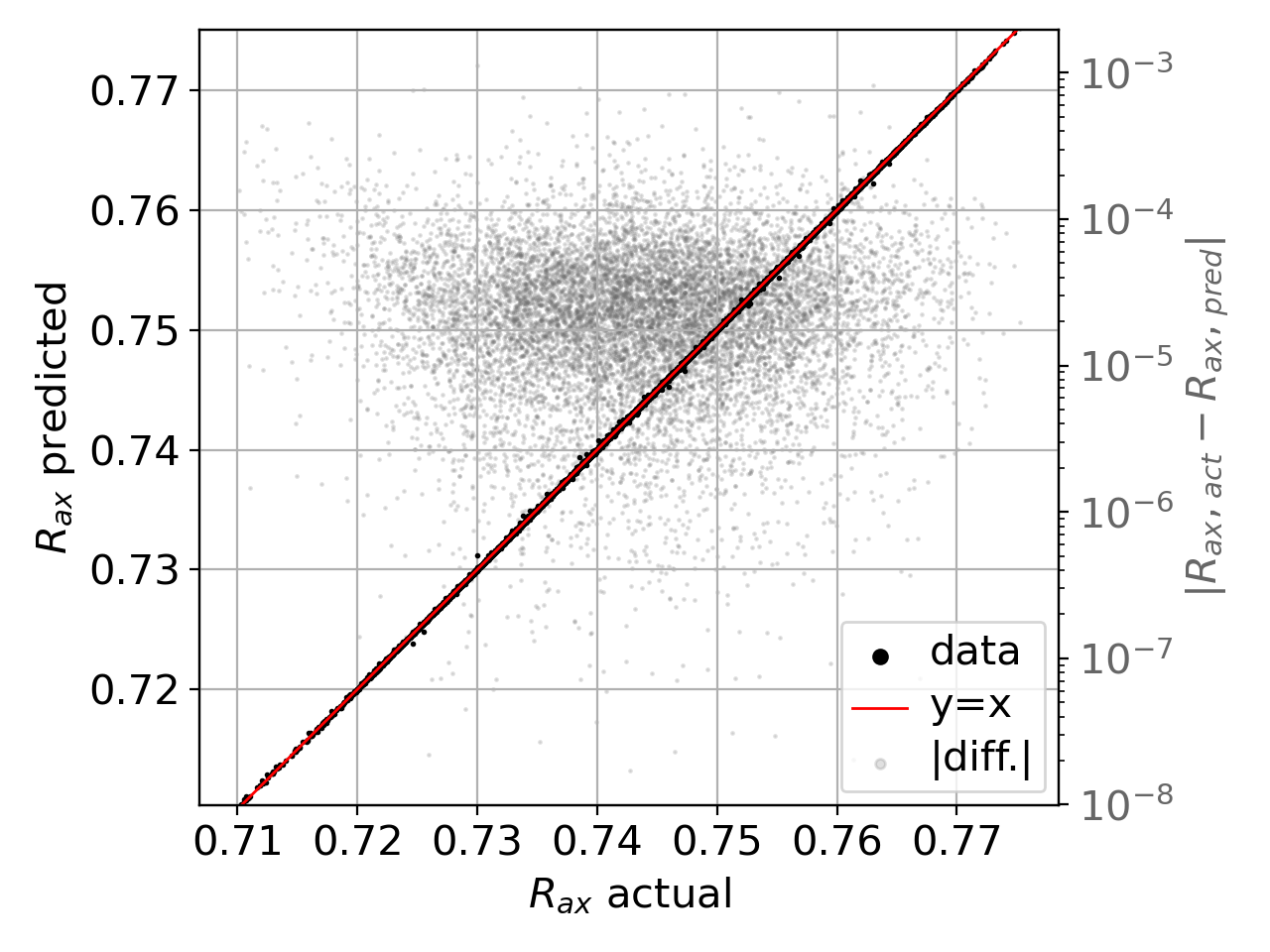}
\end{center}
\caption{$R_{ax}$ (in meters) predicted versus actual values for Test dataset (left axis) and absolute difference between them (right axis), for model shown in Table \ref{tbl:RZax_model}. $|R_{ax,act}-R_{ax,pred}|$ is distributed with median $\approx\num{2.33e-5}$ and $95^{th}$ percentile $\approx\num{1.11e-4}$.}
\label{fig:Rax_test}
\end{figure}

\begin{figure}[!htb]
\begin{center}
\includegraphics[width=\columnwidth]{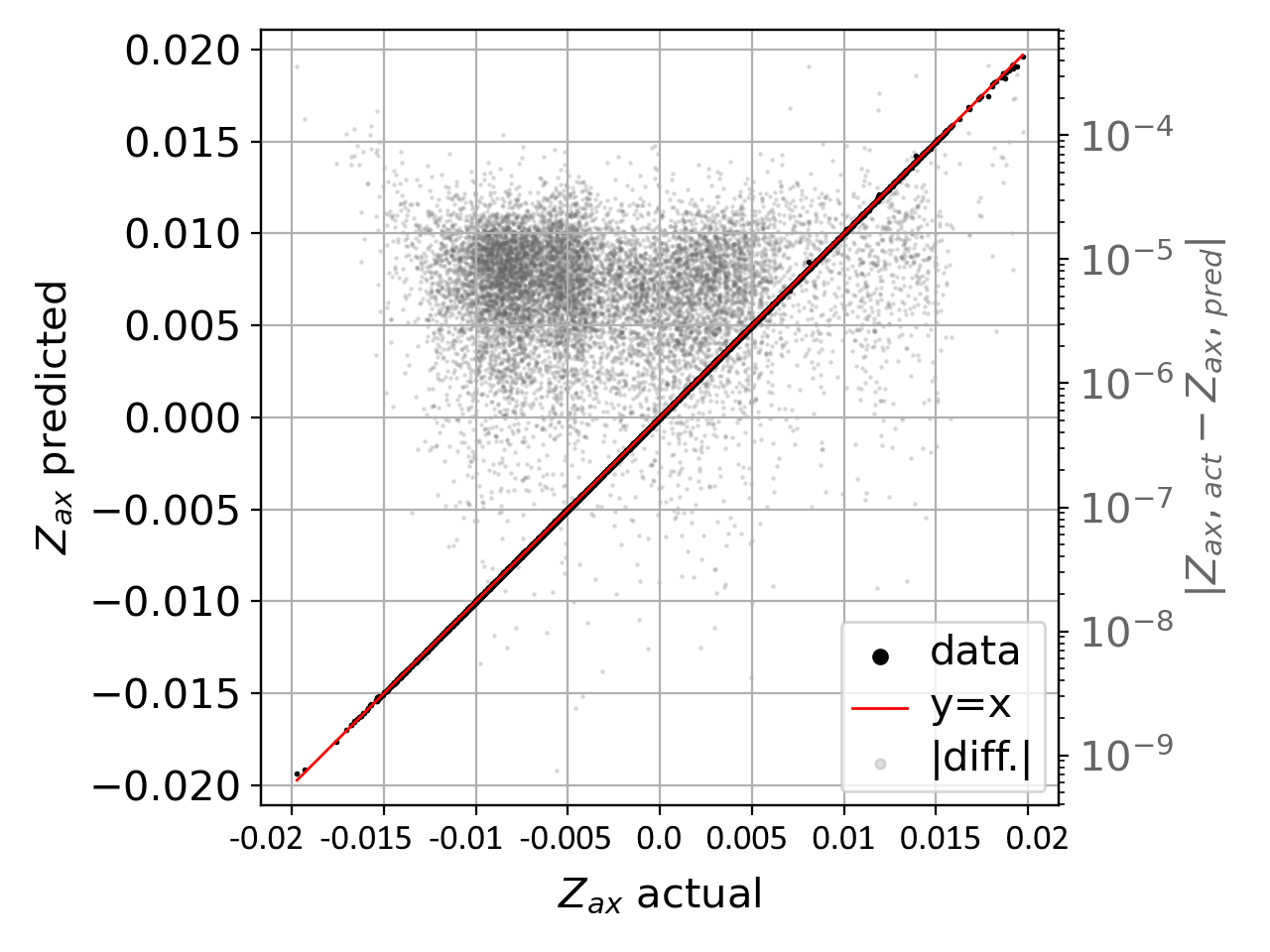}
\end{center}
\caption{$Z_{ax}$ (in meters) predicted versus actual values for Test dataset (left axis) and absolute difference between them (right axis), for model shown in Table \ref{tbl:RZax_model}. $|Z_{ax,act}-Z_{ax,pred}|$ is distributed with median $\approx\num{6.53e-6}$ and $95^{th}$ percentile $\approx\num{2.81e-5}$.}
\label{fig:Zax_test}
\end{figure}

In the first model, $R_{ax}$ and $Z_{ax}$ are predicted simultaneously. The input variables used for this model are $B_\phi$, $V_{loop}$, $I_p$, coil currents $I_{coils}$, and $B_{\theta,MC}$. After performing the \texttt{keras-tuner} hyper-optimization, the model with least validation loss is as shown in Table \ref{tbl:RZax_model}. The plot of MSE loss for training and validation datasets with respect to epochs is shown in Figure \ref{fig:RZax_loss}. Comparison and error in actual versus predicted values of $R_{ax}$ and $Z_{ax}$ for test dataset are shown in Figures \ref{fig:Rax_test} and \ref{fig:Zax_test} respectively.

{\centering
\begin{longtblr}{
    colspec  = {l l l},
    rowhead  = 0,        
    rowfoot  = 0,        
  }
\toprule
Layer & Specification & Activation \\
\midrule
Input    & 26 Units & --- \\
\midrule
Dense & 56 Units & tanh \\
Dense & 48 Units & silu \\
Dense & 36 Units & gelu \\
\midrule
Output & 2 Units & linear \\
\bottomrule
\end{longtblr}
\addtocounter{table}{-1}
\captionof{table}{Dense Model to predict $R_{ax},Z_{ax}$. Total trainable parameters: 6,086.}
\label{tbl:RZax_model}
}

\subsection{Poloidal Beta and Internal Inductance Model}
\label{sec:betap_li}
\begin{figure}[!htb]
\begin{center}
\includegraphics[width=\columnwidth]{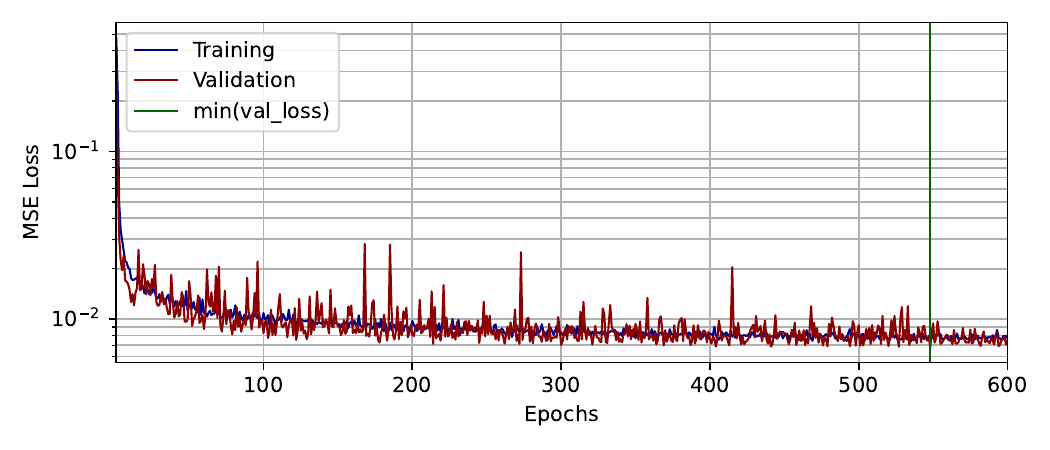}
\end{center}
\caption{MSE Loss for Training and Validation Datasets for model shown in Table \ref{tbl:betap_li_model}.}
\label{fig:betap_li_loss}
\end{figure}

\begin{figure}[!htb]
\begin{center}
\includegraphics[width=\columnwidth]{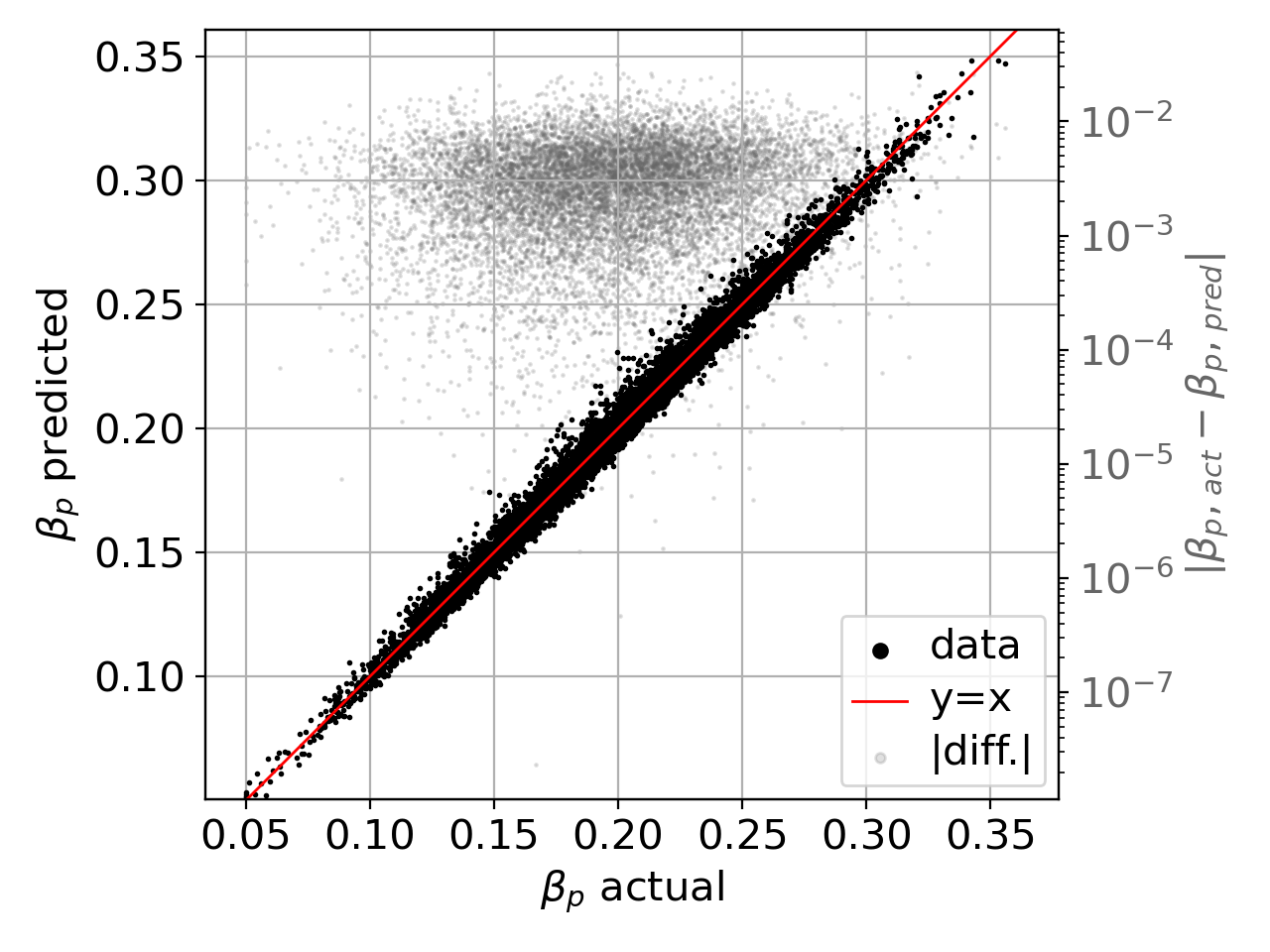}
\end{center}
\caption{$\beta_p$ predicted versus actual values for Test dataset (left axis) and absolute difference between them (right axis), for model shown in Table \ref{tbl:betap_li_model}. $|\beta_{p,act}-\beta_{p,pred}|$ is distributed with median $\approx\num{2.95e-3}$ and $95^{th}$ percentile $\approx\num{9.36e-3}$.}
\label{fig:betap_test}
\end{figure}

\begin{figure}[!htb]
\begin{center}
\includegraphics[width=\columnwidth]{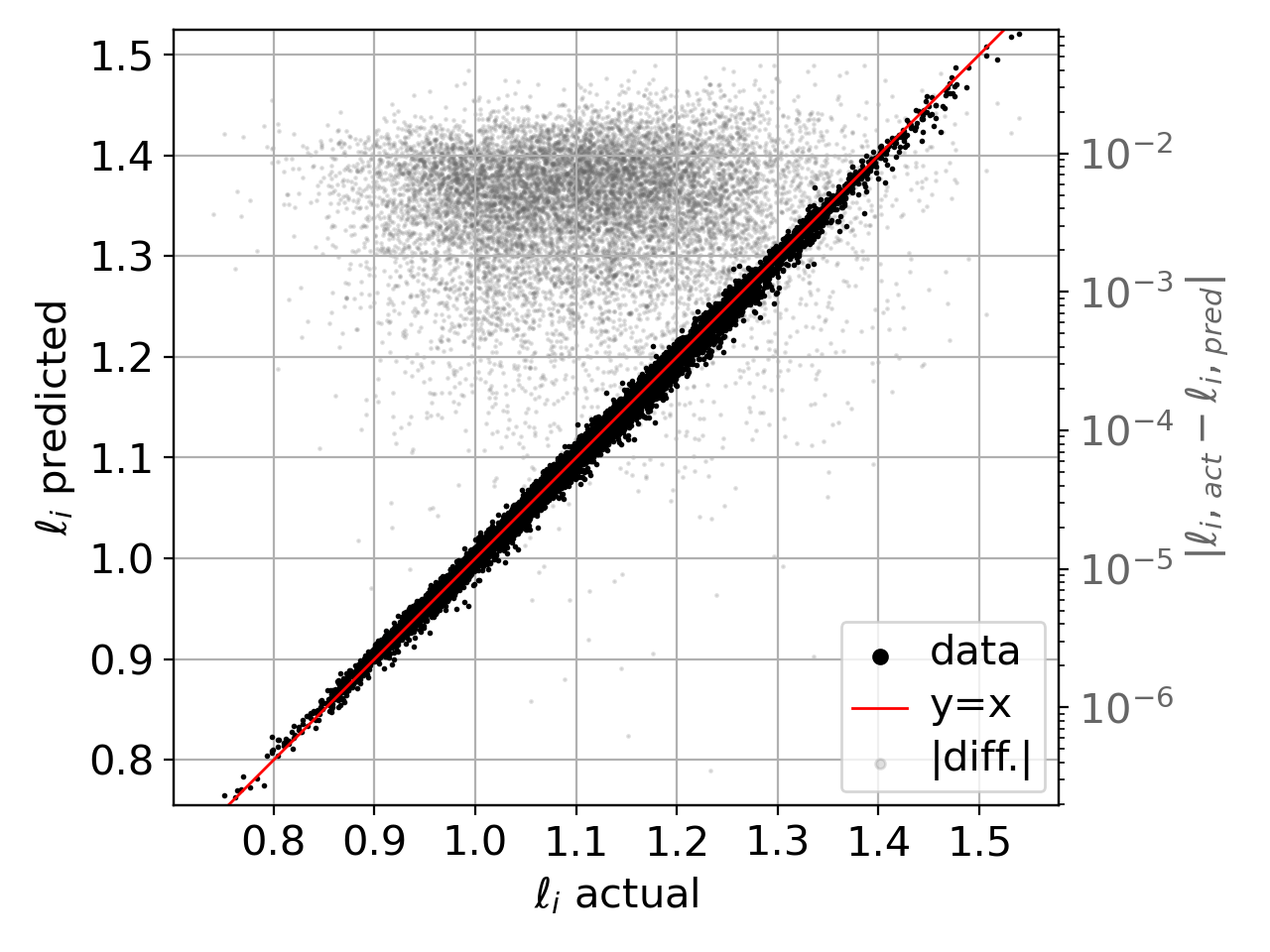}
\end{center}
\caption{$\ell_i$ predicted versus actual values for Test dataset (left axis) and absolute difference between them (right axis), for model shown in Table \ref{tbl:betap_li_model}. $|\ell_{i,act}-\ell_{i,pred}|$ is distributed with median $\approx\num{4.88e-3}$ and $95^{th}$ percentile $\approx\num{1.55e-2}$.}
\label{fig:li_test}
\end{figure}

In the second model, $\beta_p$ and $\ell_i$ are predicted simultaneously. They were also modeled separately but there was no noticeable improvement in prediction accuracy. In addition to the input variables used in the model of Section \ref{sec:RZax}, the plasma position $R_{ax}$ and $Z_{ax}$ have also been used as input variables. This is because these parameters could be predicted quite accurately as shown in Figures \ref{fig:Rax_test} and \ref{fig:Zax_test} respectively. In practice, the model to predict magnetic axis position discussed in Section \ref{sec:RZax} can be used in cascade to this model. However, during the training of this model, the dataset $R_{ax}$ and $Z_{ax}$ values were used and thus effect of cascading errors are not reflected in reported performance. After the \texttt{keras-tuner} hyper-optimization, the best-performing model found is shown in Table \ref{tbl:betap_li_model}. The plot of MSE loss for training and validation datasets with respect to epochs is shown in Figure \ref{fig:betap_li_loss}. Comparison and error in actual versus predicted values of $\beta_p$ and $\ell_i$ for test dataset are shown in Figures \ref{fig:betap_test} and \ref{fig:li_test} respectively.

{\centering
\begin{longtblr}{
    colspec  = {l l l},
    rowhead  = 0,        
    rowfoot  = 0,        
  }
\toprule
Layer & Specification & Activation \\
\midrule
Input    & 28 Units & --- \\
\midrule
Dense & 64 Units & silu \\
Dense & 192 Units & gelu \\
Dense & 96 Units & silu \\
\midrule
Output & 2 Units & linear \\
\bottomrule
\end{longtblr}
\addtocounter{table}{-1}
\captionof{table}{Dense Model to predict $\beta_p,\ell_i$. Total trainable parameters: 33,122.}
\label{tbl:betap_li_model}
}

\subsection{Edge Safety Factor Model}
\label{sec:q1}
\begin{figure}[!htb]
\begin{center}
\includegraphics[width=\columnwidth]{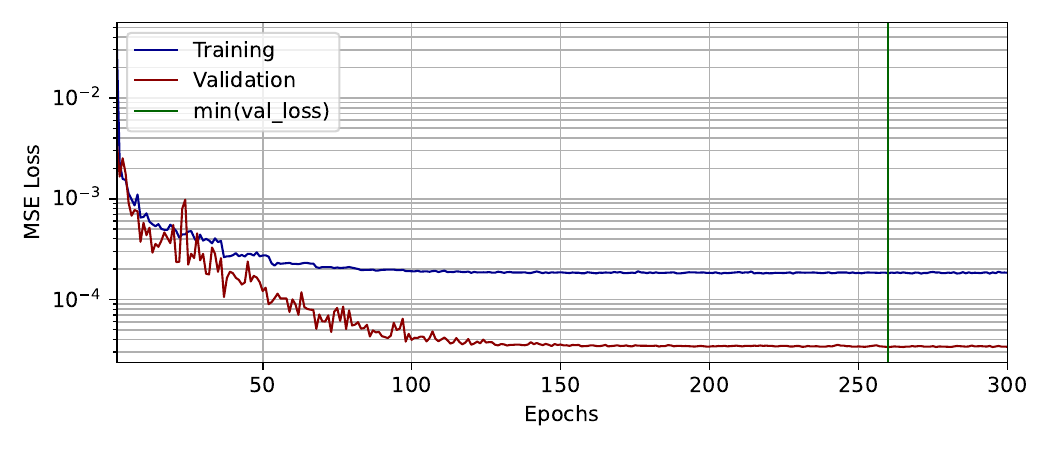}
\end{center}
\caption{MSE Loss for Training and Validation Datasets for model shown in Table \ref{tbl:q1_model}.}
\label{fig:q1_loss}
\end{figure}

\begin{figure}[!htb]
\begin{center}
\includegraphics[width=\columnwidth]{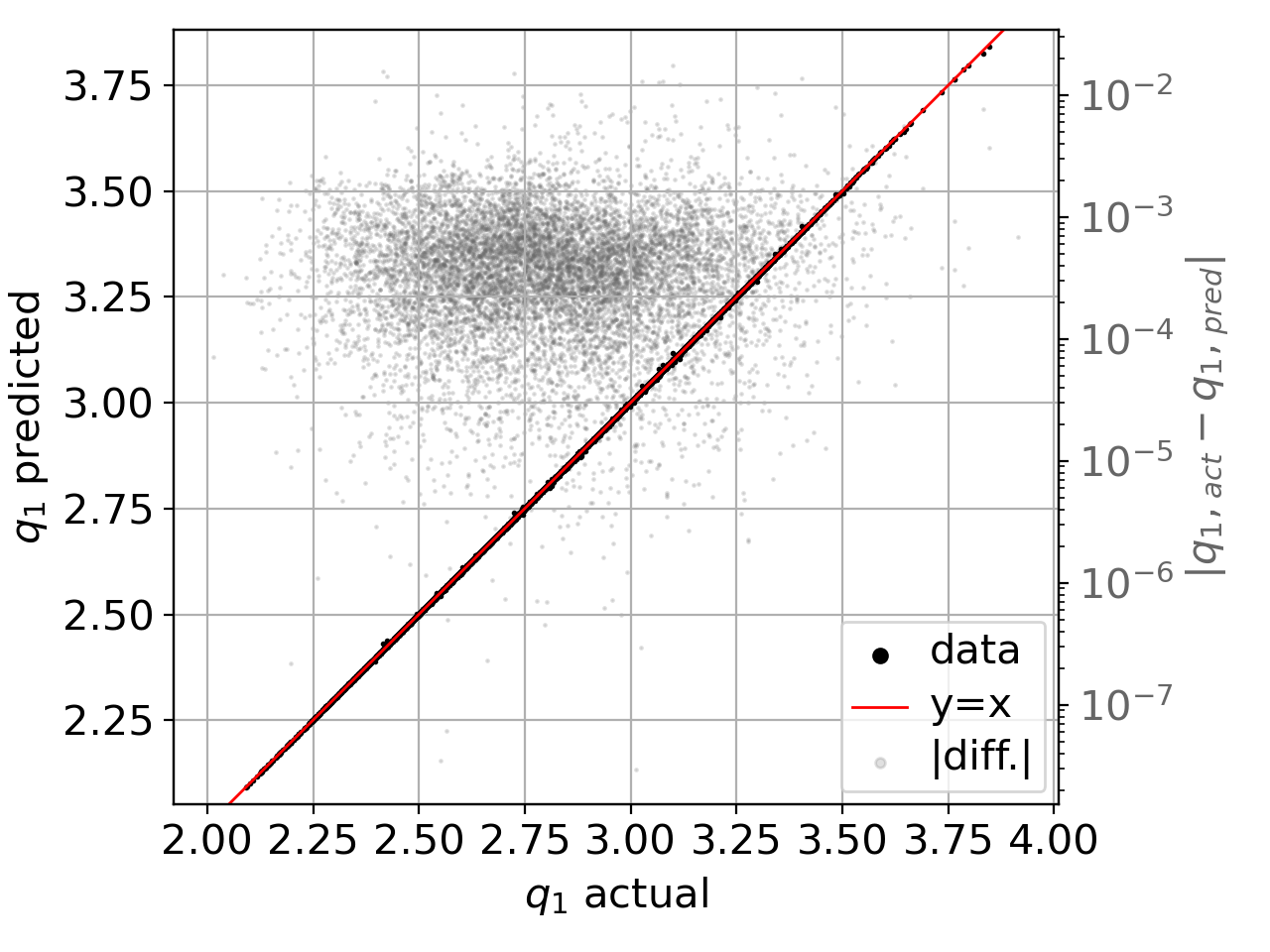}
\end{center}
\caption{$q_1$ predicted versus actual values for Test dataset (left axis) and absolute difference between them (right axis), for model shown in Table \ref{tbl:q1_model}. $|q_{1,act}-q_{1,pred}|$ is distributed with median $\approx\num{3.9e-4}$ and $95^{th}$ percentile $\approx\num{1.79e-3}$.}
\label{fig:q1_test}
\end{figure}

In this model, edge safety factor $q_1$ is predicted. Even though full $q$ profile prediction models are discussed in Section \ref{sec:q}, $q_1$ was modeled separately so that it can be used as an input for $\psi$ models discussed in Section \ref{sec:psi}. The input variables used in this model are same as inputs used in the model of $\beta_p$ and $\ell_i$ in Section \ref{sec:betap_li}. After the \texttt{keras-tuner} hyper-optimization, the best-performing model found is shown in Table \ref{tbl:q1_model}. The plot of MSE loss for training and validation datasets with respect to epochs is shown in Figure \ref{fig:q1_loss}. The validation loss is mostly lower than training loss because dropout regularization is active during training but disabled during validation. The use of Dropout layer was necessary in this model, without which significant overfitting was observed. Comparison and error in actual versus predicted values of $q_1$ for test dataset are shown in Figure \ref{fig:q1_test}.

{\centering
\begin{longtblr}{
    colspec  = {l l l},
    rowhead  = 0,        
    rowfoot  = 0,        
  }
\toprule
Layer & Specification & Activation \\
\midrule
Input    & 28 Units & --- \\
\midrule
Dense & 224 Units & silu \\
Dense & 96 Units & relu \\
Dense & 128 Units & gelu \\
Dropout & Rate 0.02 & --- \\
\midrule
Output & 1 Unit & linear \\
\bottomrule
\end{longtblr}
\addtocounter{table}{-1}
\captionof{table}{Dense Model to predict $q_1$. Total trainable parameters: 40,641.}
\label{tbl:q1_model}
}

\subsection{Limiter and Axis $\psi$ Model}
\label{sec:psi01}
\begin{figure}[!htb]
\begin{center}
\includegraphics[width=\columnwidth]{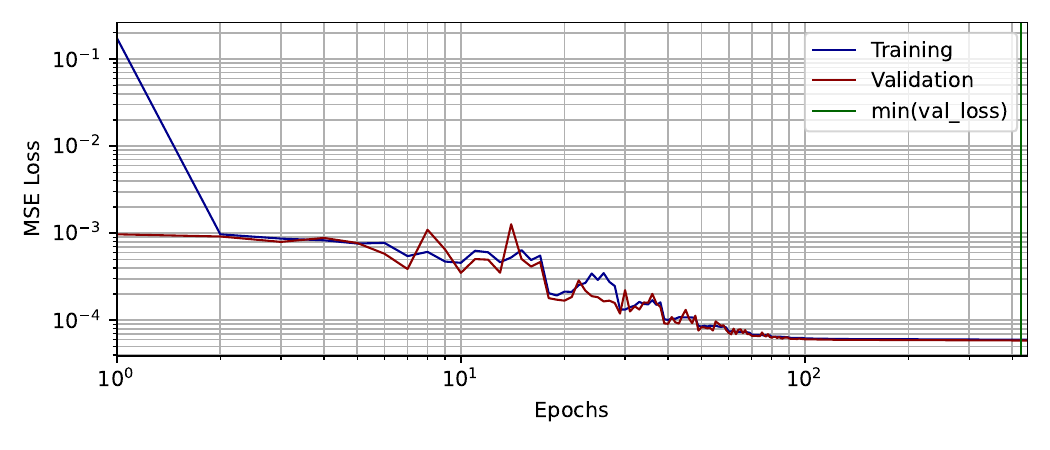}
\end{center}
\caption{MSE Loss for Training and Validation Datasets for model shown in Table \ref{tbl:psi01_model}.}
\label{fig:psi01_loss}
\end{figure}

\begin{figure}[!htb]
\begin{center}
\includegraphics[width=\columnwidth]{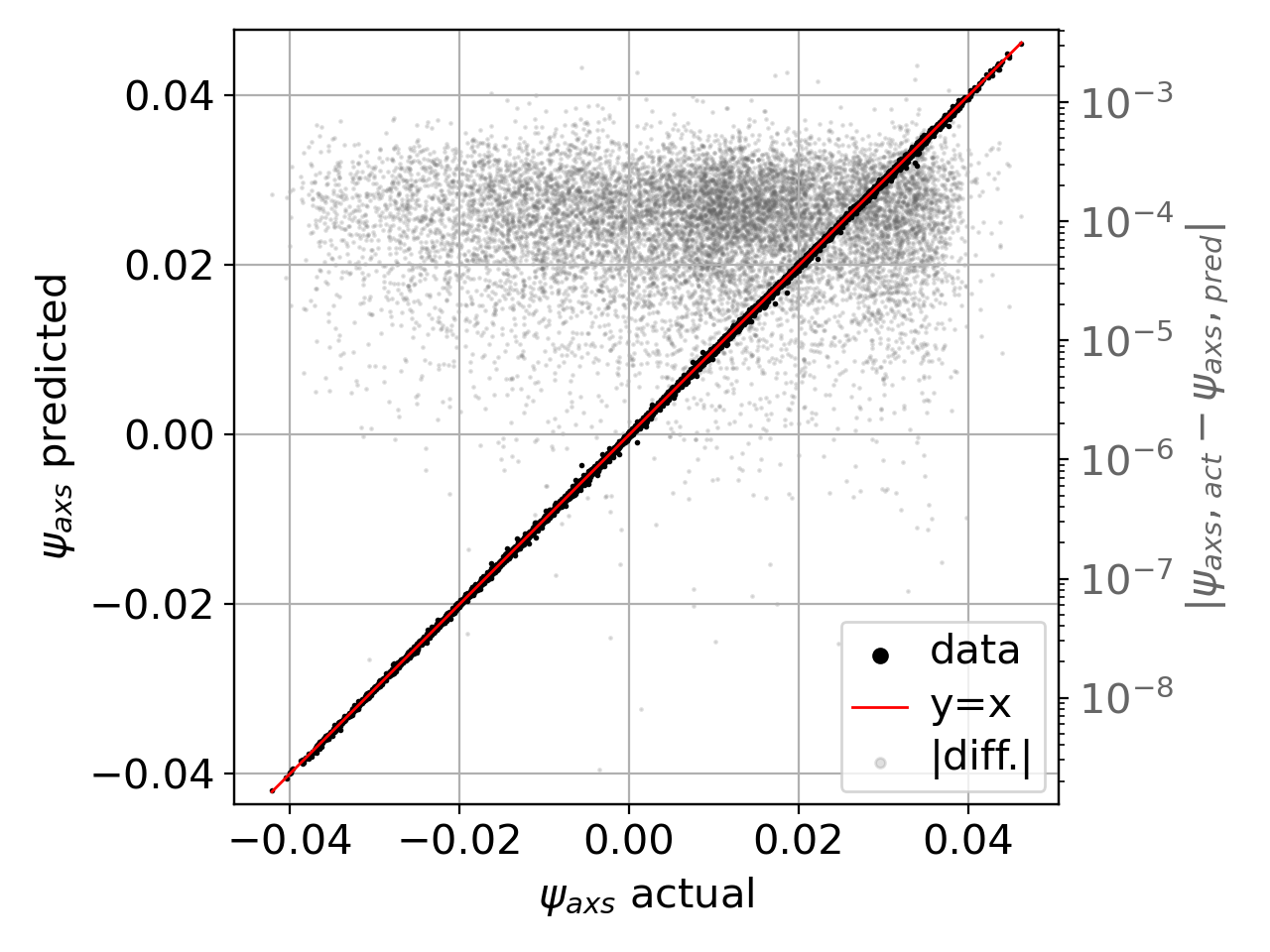}
\end{center}
\caption{$\psi_{axs}$ predicted versus actual values for Test dataset (left axis) and absolute difference between them (right axis), for model shown in Table \ref{tbl:psi01_model}. $|\psi_{axs,act}-\psi_{axs,pred}|$ is distributed with median $\approx\num{1.06e-4}$ and $95^{th}$ percentile $\approx\num{3.67e-4}$.}
\label{fig:psiaxs_test}
\end{figure}

\begin{figure}[!htb]
\begin{center}
\includegraphics[width=\columnwidth]{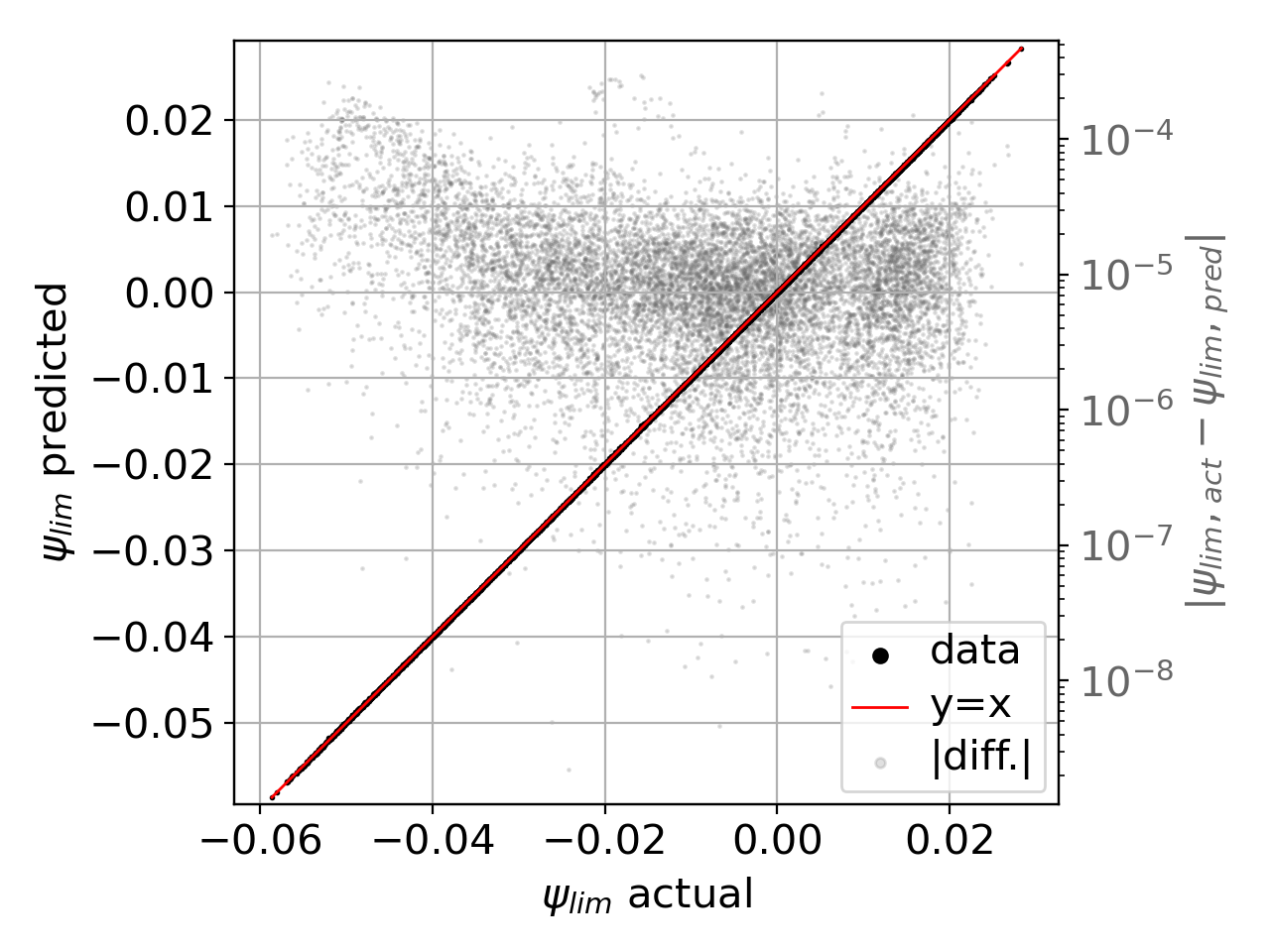}
\end{center}
\caption{$\psi_{lim}$ predicted versus actual values for Test dataset (left axis) and absolute difference between them (right axis), for model shown in Table \ref{tbl:psi01_model}. $|\psi_{lim,act}-\psi_{lim,pred}|$ is distributed with median $\approx\num{8.47e-6}$ and $95^{th}$ percentile $\approx\num{4.60e-5}$.}
\label{fig:psilim_test}
\end{figure}

In this model, $\psi$ values at magnetic axis $\psi_{axs}$ and at limiter $\psi_{lim}$ are predicted. Models regarding full $\psi$ profile are discussed in Section \ref{sec:psi}, but these two are modeled separately so that they can be used as an inputs for full $\psi$ models. The input variables used in this model are same as inputs used in the model of $\beta_p$ and $\ell_i$ in Section \ref{sec:betap_li}. After the \texttt{keras-tuner} hyper-optimization, the best-performing model found is shown in Table \ref{tbl:psi01_model}. The plot of MSE loss for training and validation datasets with respect to epochs is shown in Figure \ref{fig:psi01_loss}. Comparison and error in actual versus predicted values of $\psi_{axs}$ and $\psi_{lim}$ for test dataset are shown in Figures \ref{fig:psiaxs_test} and \ref{fig:psilim_test}.

{\centering
\begin{longtblr}{
    colspec  = {l l l},
    rowhead  = 0,        
    rowfoot  = 0,        
  }
\toprule
Layer & Specification & Activation \\
\midrule
Input    & 28 Units & --- \\
\midrule
Dense & 224 Units & gelu \\
Dense & 224 Units & gelu \\
Dense & 224 Units & softplus \\
Dense & 224 Units & softplus \\
\midrule
Output & 2 Units & linear \\
\bottomrule
\end{longtblr}
\addtocounter{table}{-1}
\captionof{table}{Dense Model to predict $\psi_{axs}$ and $\psi_{lim}$. Total trainable parameters: 154,786.}
\label{tbl:psi01_model}
}

\subsection{Coil Currents Inverse Model}
\label{sec:cc}
\begin{figure}[!htb]
\begin{center}
\includegraphics[width=\columnwidth]{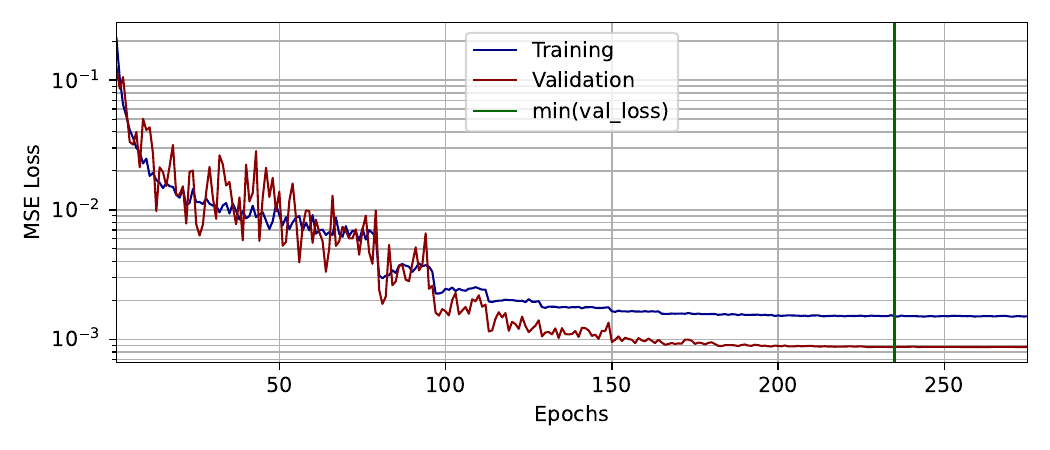}
\end{center}
\caption{MSE Loss for Training and Validation Datasets for model shown in Table \ref{tbl:cc_model}.}
\label{fig:cc_loss}
\end{figure}

\begin{figure}[!htb]
\begin{center}
\includegraphics[width=\columnwidth]{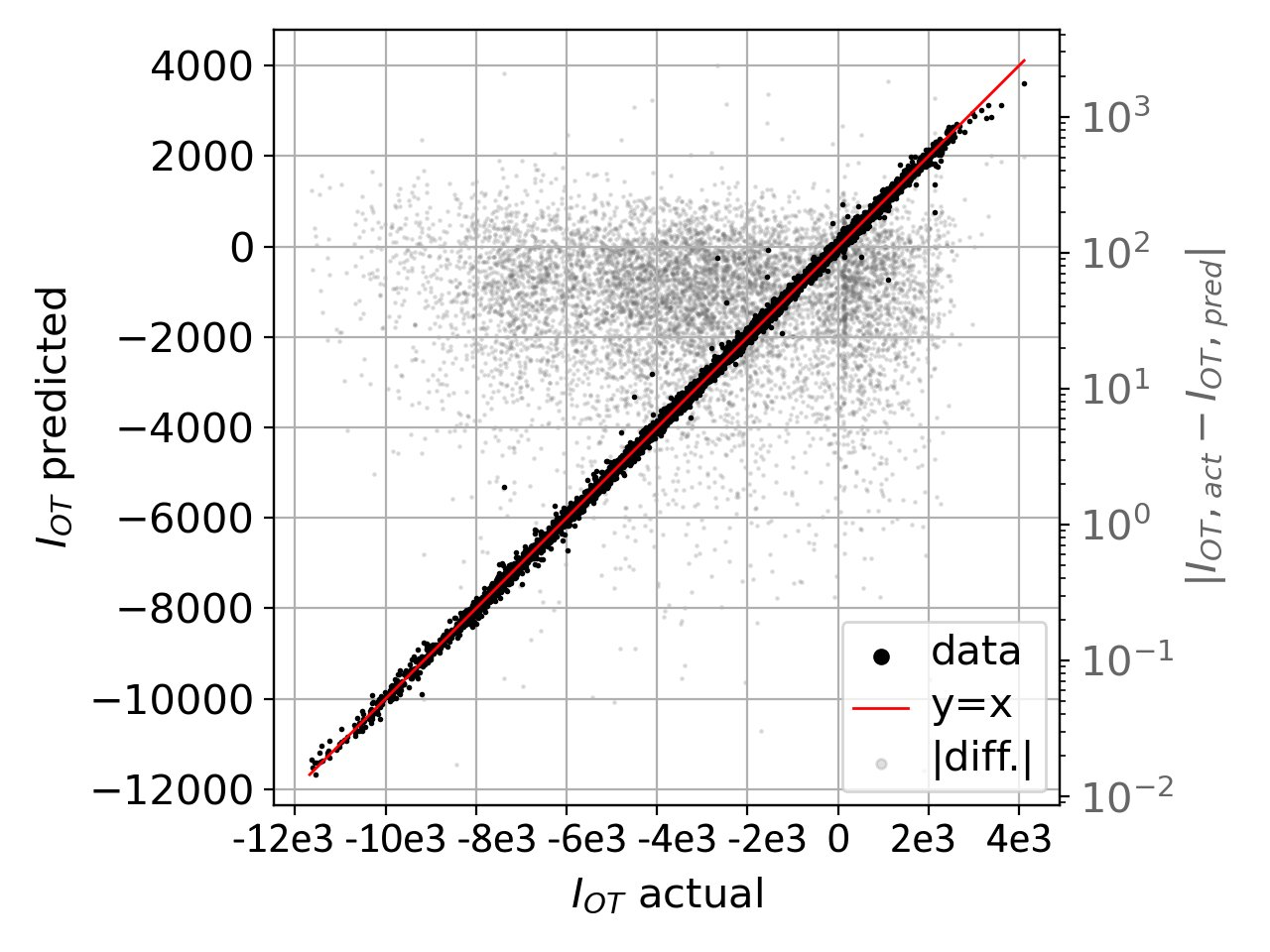}
\end{center}
\caption{$I_{OT}$ (in Ampere/turn) predicted versus actual values for Test dataset (left axis) and absolute difference between them (right axis), for model shown in Table \ref{tbl:cc_model}. $|I_{OT,act}-I_{OT,pred}|$ is distributed with median $\approx\num{67.9}$ and $95^{th}$ percentile $\approx\num{260}$.}
\label{fig:IOT_test}
\end{figure}

\begin{figure}[!htb]
\begin{center}
\includegraphics[width=\columnwidth]{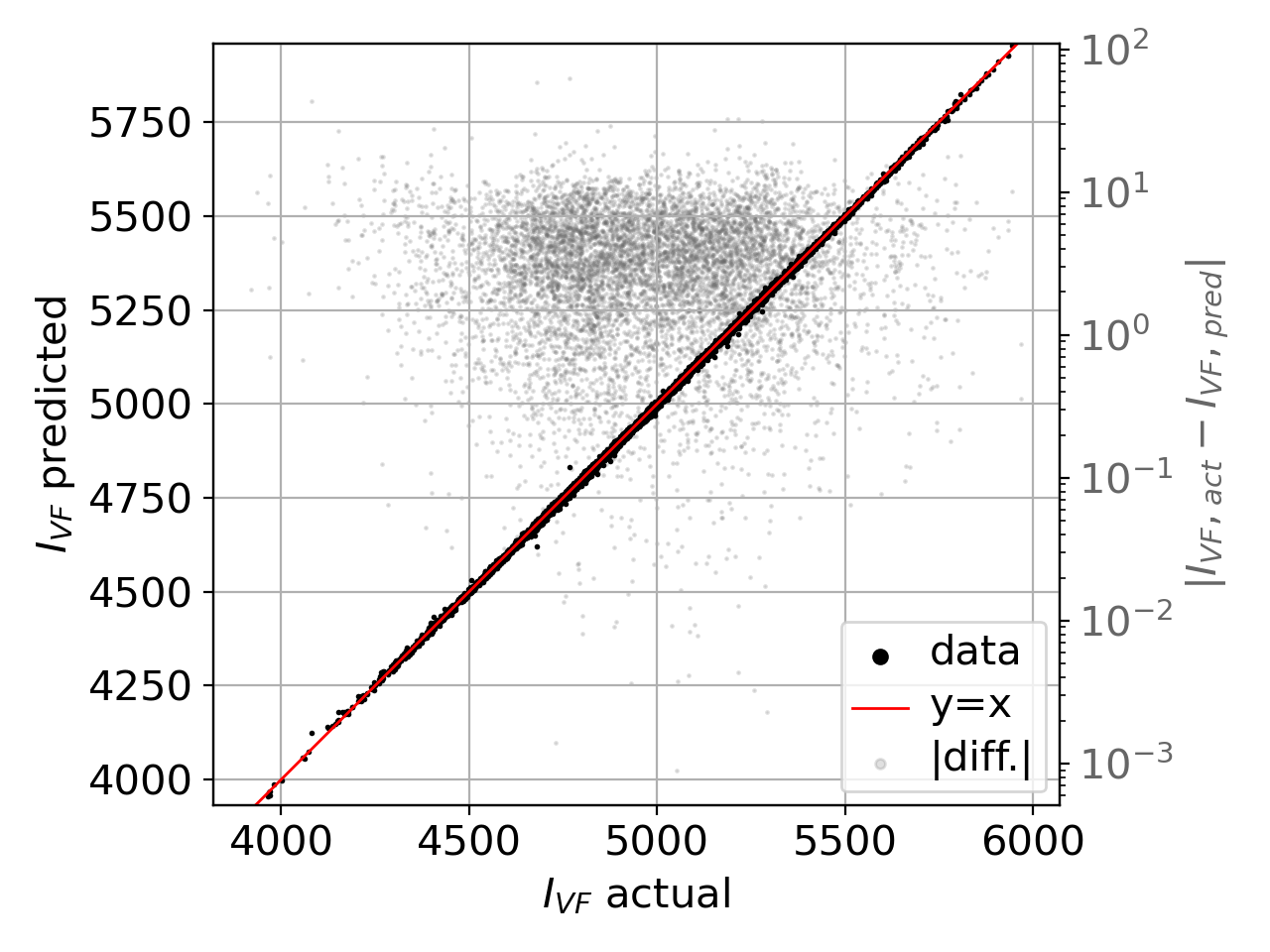}
\end{center}
\caption{$I_{VF}$ (in Ampere/turn) predicted versus actual values for Test dataset (left axis) and absolute difference between them (right axis), for model shown in Table \ref{tbl:cc_model}. $|I_{VF,act}-I_{VF,pred}|$ is distributed with median $\approx\num{3.22}$ and $95^{th}$ percentile $\approx\num{10.0}$.}
\label{fig:IVF_test}
\end{figure}

\begin{figure}[!htb]
\begin{center}
\includegraphics[width=\columnwidth]{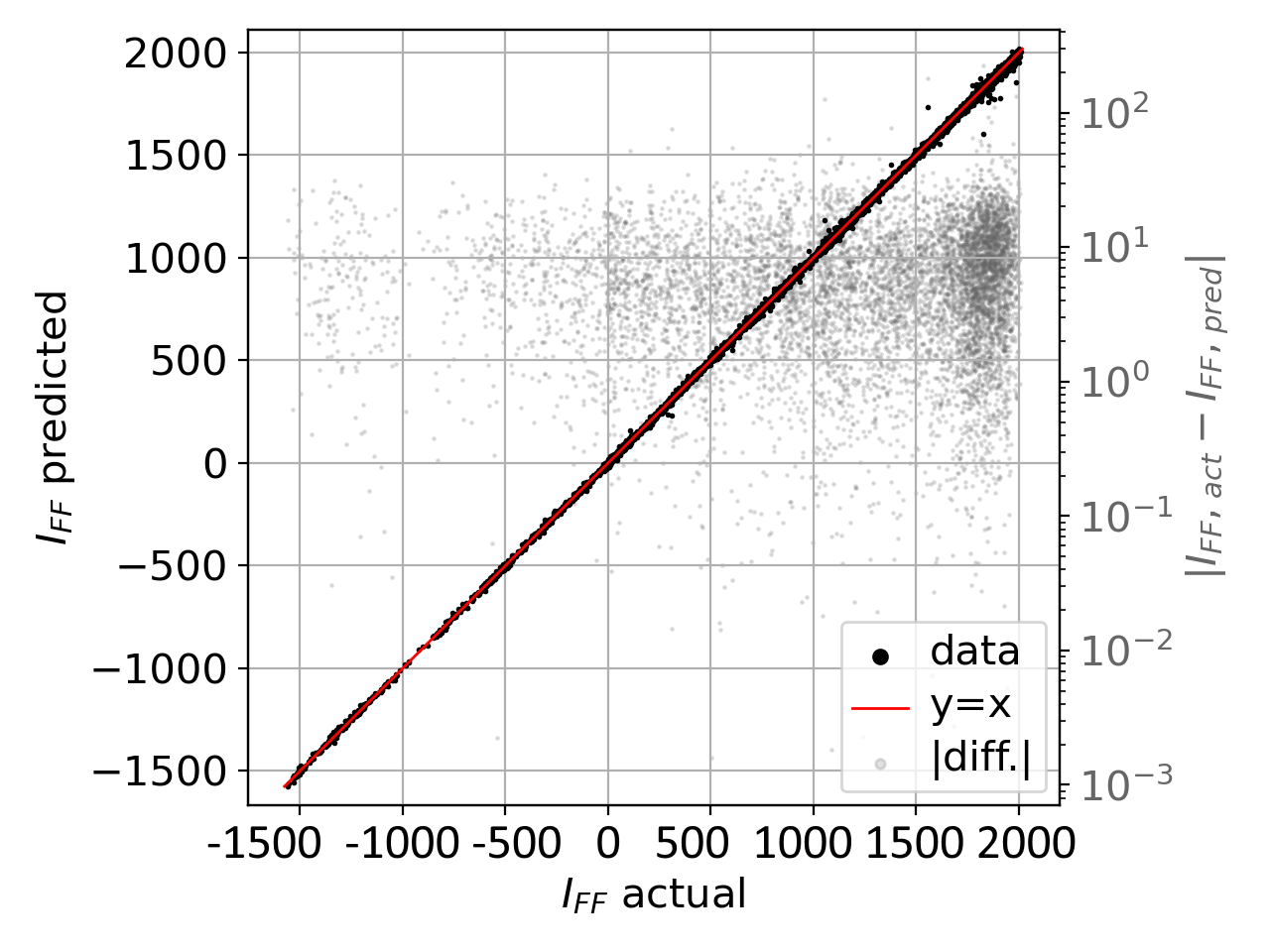}
\end{center}
\caption{$I_{FF}$ (in Ampere/turn) predicted versus actual values for Test dataset (left axis) and absolute difference between them (right axis), for model shown in Table \ref{tbl:cc_model}. $|I_{FF,act}-I_{FF,pred}|$ is distributed with median $\approx\num{6.24}$ and $95^{th}$ percentile $\approx\num{23.2}$.}
\label{fig:IFF_test}
\end{figure}

In this inverse model, poloidal field coil currents are predicted from desired equilibrium parameters. The input variables used in this model are $B_\phi$, $R_{ax}$, $Z_{ax}$, $I_p$, $\beta_p$, $q_0$, $q_1$, and $\ell_i$. The output variables are $I_{OT}$, $I_{VF}$, and $I_{FF}$. Prediction of divertor coil currents was also attempted but resulted in poor performance because only a small fraction of the dataset contains nonzero divertor coil currents and their magnitudes are relatively small. After performing the \texttt{keras-tuner} hyper-optimization, the model with least validation loss is shown in Table \ref{tbl:cc_model}. The plot of MSE loss for training and validation datasets with respect to epochs is shown in Figure \ref{fig:cc_loss}. Similar to the $q_1$ model discussed in Section \ref{sec:q1}, the validation loss is occasionally lower than the training loss because dropout regularization is active during training but disabled during validation.  Comparison and error in actual versus predicted values of coil currents for test dataset are shown in Figures \ref{fig:IOT_test}, \ref{fig:IVF_test} and \ref{fig:IFF_test}. The mapping from input equilibrium parameters to coil currents is not unique and thus the predicted currents should not be interpreted as unique actuator settings. Instead, the model learns statistically preferred operational relationships and predicts coil-current combinations represented within the dataset. Nonetheless, the predictive performance indicates that the model should be useful for experiment planning and preliminary actuator selection, within the operational space of the dataset.

{\centering
\begin{longtblr}{
    colspec  = {l l l},
    rowhead  = 0,        
    rowfoot  = 0,        
  }
\toprule
Layer & Specification & Activation \\
\midrule
Input    & 8 Units & --- \\
\midrule
Dense & 160 Units & silu \\
Dense & 32 Units & selu \\
Dense & 80 Units & silu \\
Dense & 208 Units & silu \\
Dropout & Rate 0.04 & --- \\
\midrule
Output & 3 Units & linear \\
\bottomrule
\end{longtblr}
\addtocounter{table}{-1}
\captionof{table}{Dense Model to predict coil currents $I_{OT}$, $I_{VF}$ and $I_{FF}$. Total trainable parameters: 26,707.}
\label{tbl:cc_model}
}

\section{Safety Factor Profile Models}
\label{sec:q}
The equilibrium dataset contains safety factor $q$ profile for each case with respect to $\rho=1-\bar\psi$ as one-dimensional array of size 101. The $q$ profiles in the dataset exhibit a relatively restricted family of shapes, characterized by smooth and predominantly monotonic variation from the magnetic axis to the plasma edge as shown in Figure \ref{fig:q_dist}. This behavior suggests that the profiles may be represented efficiently using a low-dimensional basis obtained through PCA. Alternatively, the local correlations between neighboring radial locations make 1d-CNN a suitable choice for direct profile prediction. Besides providing alternative prediction strategies, developing these models provides a direct comparison between dimensionality-reduction based reconstruction and local spatial feature learning. PCA based modeling and results are discussed in Section \ref{sec:q_pca} and 1d-CNN based modeling and results are discussed in Section \ref{sec:q_cnn}. Both models use the same input variables as the $\beta_p$ and $\ell_i$ model discussed in Section \ref{sec:betap_li}.

\subsection{PCA Model for $q$-profile}
\label{sec:q_pca}
\begin{figure}[!htb]
\begin{center}
\includegraphics[width=\columnwidth]{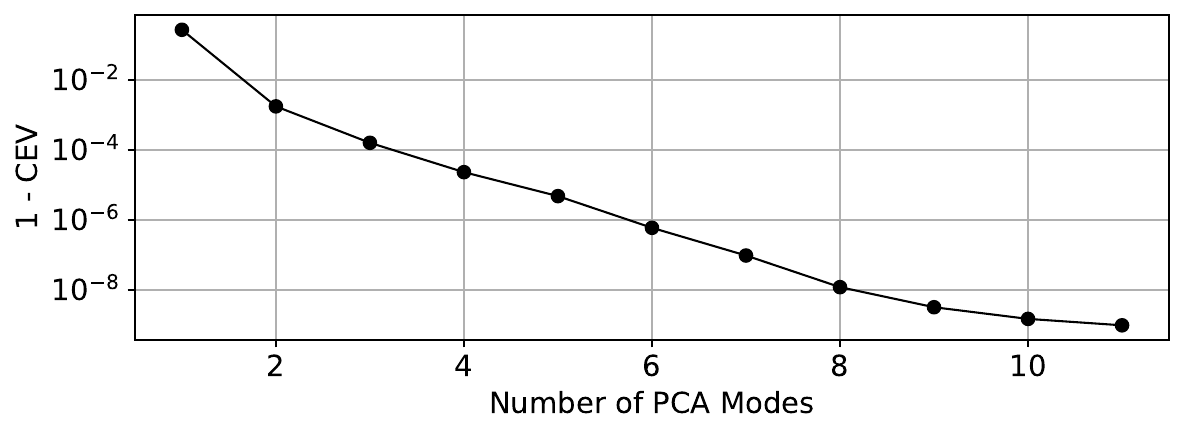}
\end{center}
\caption{$1-CEV$ with respect to number of retained PCA modes for the $q$-profiles.}
\label{fig:q_pca_cev}
\end{figure}

\begin{figure}[!htb]
\begin{center}
\includegraphics[width=\columnwidth]{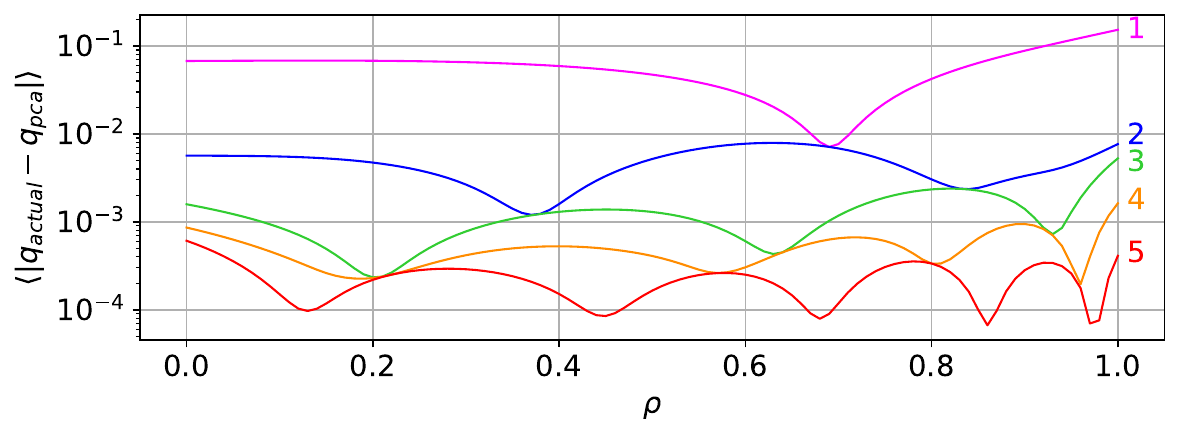}
\end{center}
\caption{Mean absolute difference between actual and reconstructed $q$-profile for different number of retained PCA modes. The colored numbers on right side denote the numbers of PCA modes.}
\label{fig:q_pca_err}
\end{figure}

To determine the number of PCA modes required for the dataset of $q$-profile, the Cumulative Explained Variance (CEV) was checked for different number of retained PCA modes. The CEV quantifies the fraction of total dataset variance captured by a given number of PCA modes. Although CEV is a useful metric for selecting modes in smooth profile data, it may not fully characterize the reconstruction of localized profile features. For smooth profile datasets such as the present one, a CEV of 99.99\% generally indicates that the dominant profile variability has been retained. Figure \ref{fig:q_pca_cev} shows $1-CEV$ against the number of PCA modes and demonstrates that 99.99\% CEV is achieved in just 4 modes. This indicates that most of the profile variability can be represented using only 4 PCA coefficients, reducing the output dimensionality from 101 to 4, subsequently improving training stability and generalization. Figure \ref{fig:q_pca_err} shows the $\langle|q_{actual}-q_{pca}|\rangle$ for the entire dataset, which suggests that for 4 PCA modes, the mean absolute difference in reconstructed $q$-profile is mostly below \num{1e-3}, an error less than about 0.1\%.

\begin{figure}[!htb]
\begin{center}
\includegraphics[width=\columnwidth]{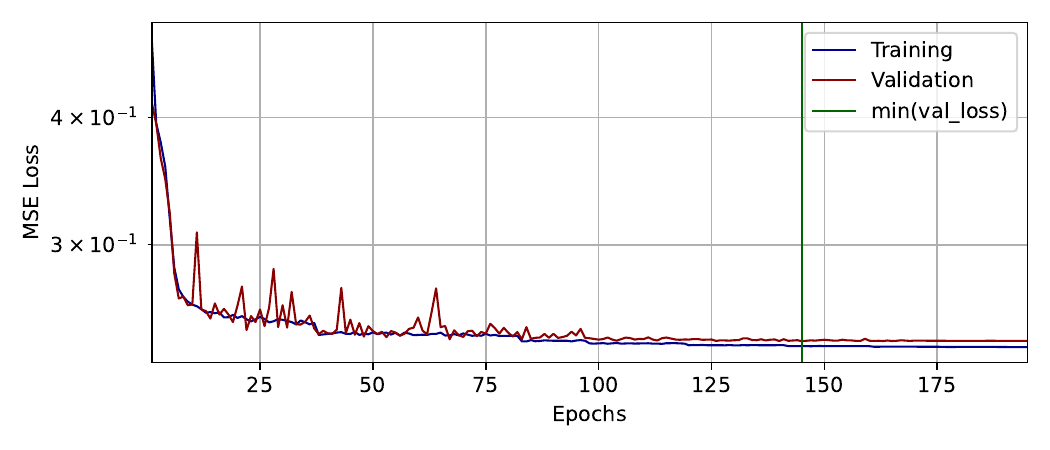}
\end{center}
\caption{MSE Loss for Training and Validation Datasets for model shown in Table \ref{tbl:q_pca_model}.}
\label{fig:q_pca_loss}
\end{figure}

The 4 retained PCA modes were modeled similar to previous Dense neural-network models discussed in Section \ref{sec:0D_Models}. After performing the \texttt{keras-tuner} hyper-optimization, the best-performing model obtained is as shown in Table \ref{tbl:q_pca_model}. The plot of MSE loss for training and validation datasets with respect to epochs is shown in Figure \ref{fig:q_pca_loss}.

{\centering
\begin{longtblr}{
    colspec  = {l l l},
    rowhead  = 0,        
    rowfoot  = 0,        
  }
\toprule
Layer & Specification & Activation \\
\midrule
Input    & 28 Units & --- \\
\midrule
Dense & 256 Units & softplus \\
Dense & 192 Units & tanh \\
Dense & 160 Units & elu \\
Dense & 64 Units & tanh \\
\midrule
Output & 4 Units & linear \\
\bottomrule
\end{longtblr}
\addtocounter{table}{-1}
\captionof{table}{Dense Model to predict 4 PCA modes of $q$-profile. Total trainable parameters: 98,212.}
\label{tbl:q_pca_model}
}

\begin{figure}[!htb]
\begin{center}
\includegraphics[width=\columnwidth]{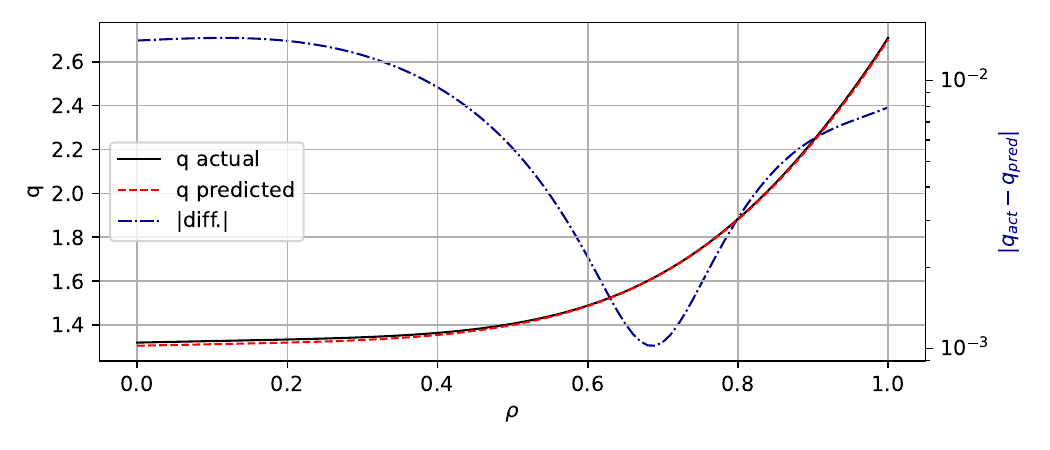}
\end{center}
\caption{Predicted $q$-profile with median MSE, for model shown in Table \ref{tbl:q_pca_model}. Values on left axis and absolute difference on right axis.}
\label{fig:q_pca_median}
\end{figure}

\begin{figure}[!htb]
\begin{center}
\includegraphics[width=\columnwidth]{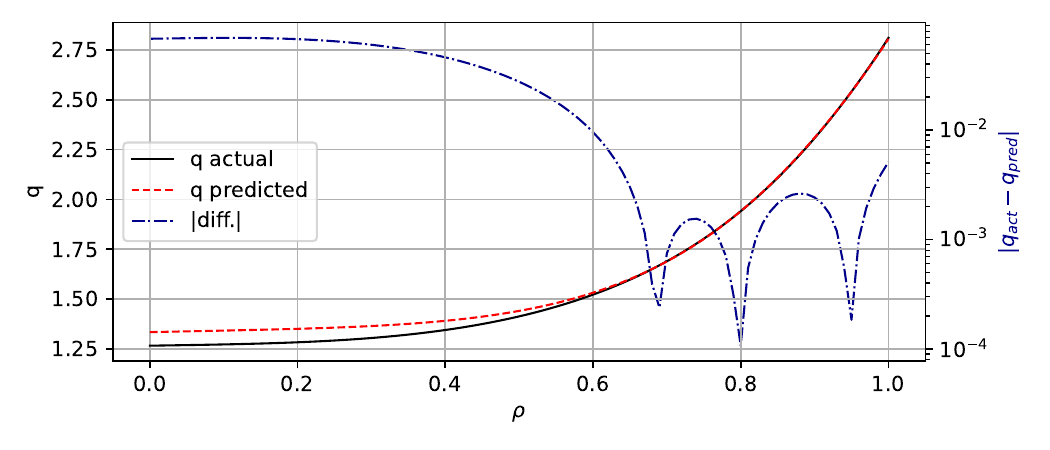}
\end{center}
\caption{Predicted $q$-profile with $99^{th}$ percentile in MSE, for model shown in Table \ref{tbl:q_pca_model}. Values on left axis and absolute difference on right axis.}
\label{fig:q_pca_99p}
\end{figure}

\begin{figure}[!htb]
\begin{center}
\includegraphics[width=\columnwidth]{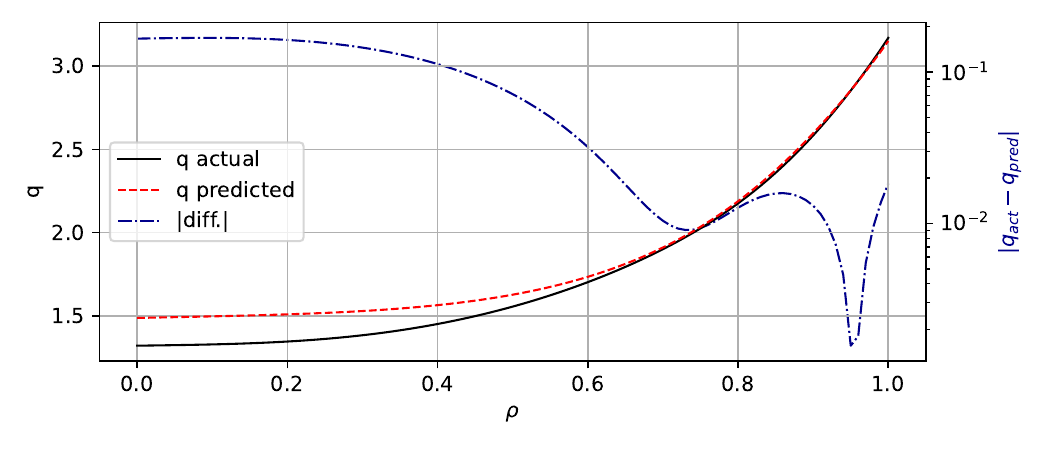}
\end{center}
\caption{Predicted $q$-profile with highest MSE, for model shown in Table \ref{tbl:q_pca_model}. Values on left axis and absolute difference on right axis.}
\label{fig:q_pca_worst}
\end{figure}

The predictive performance of the model on the test dataset is summarized in Table \ref{tbl:q_pca_performance}, which reports the $95^{th}$ and $99^{th}$ percentiles of the absolute prediction error for the PCA coefficients and for $q$ values at selected $\rho$ points. Mentioning relative errors could be misleading because some PCA coefficients have values close to zero. Therefore, only absolute errors are reported together with the corresponding variable ranges. The results show that the first two PCA coefficients are generally predicted with higher accuracy than the third and fourth coefficients,  which is expected because higher modes generally represent lower-variance profile features. Despite the larger uncertainties in the higher-order coefficients, the predicted $q$ at axis exhibits an error less than 10\% for most cases and $q$ at edge mostly has errors below 2\%.  When test dataset predictions are sorted by MSE, the median, $99^{th}$ percentile and worst-case predictions are shown in Figures \ref{fig:q_pca_median}, \ref{fig:q_pca_99p} and \ref{fig:q_pca_worst}, respectively. Even for the worst prediction in the test dataset, the error in the core safety factor is approximately 15\%, while the edge region remains well reproduced. The largest discrepancies generally occur near the magnetic axis, where the profile is more sensitive to details of the $J_\phi$ distribution that are not directly represented in the model inputs.

{\centering
\begin{longtblr}{
    colspec  = {p{0.15\linewidth} p{0.12\linewidth} p{0.26\linewidth} p{0.26\linewidth}},
    rowhead  = 0,        
    rowfoot  = 0,        
  }
\toprule
Variable & Data Range & \footnotesize{$95^{th}$ percentile in Absolute Error} & \footnotesize{$99^{th}$ percentile in Absolute Error} \\
\midrule
PCA Mode 1 & \small{-3.99, 4.90} & 0.187 & 0.283 \\
PCA Mode 2 & \small{-2.65, 2.65} & 0.190 & 0.294 \\
PCA Mode 3 & \small{-0.157, 0.445} & 0.107 & 0.161 \\
PCA Mode 4 & \footnotesize{-0.0660, 0.138} & 0.0153 & 0.0256 \\
\midrule
$q$($\rho$=0) & \small{0.899, 1.50} & 0.0442 & 0.0683 \\
$q$($\rho$=0.1) & \small{0.90, 1.51} & 0.0448 & 0.0690 \\
$q$($\rho$=0.9) & \small{1.63, 3.08} & 0.0134 & 0.0214 \\
$q$($\rho$=1) & \small{2.00, 4.00} & 0.0173 & 0.0287 \\
\bottomrule
\end{longtblr}
\addtocounter{table}{-1}
\captionof{table}{Performance in predictions of 4 PCA modes and $q$ at some $\rho$ values, for test dataset using the model shown in Table \ref{tbl:q_pca_model}.}
\label{tbl:q_pca_performance}
}

\subsection{1d-CNN Model for $q$-profile}
\label{sec:q_cnn}
In this approach, $q$-profiles are modeled using 1d-CNN. Rather than predicting the $q$ profile directly, the model predicts $q'=\partial q/\partial\rho$. This formulation exploits the fact that $q$ increases monotonically with $\rho$ in the generated equilibrium dataset. Consequently, $q'$ is expected to remain positive for all $\rho$, enabling the use of the \texttt{softplus} activation function in the output layer. This enforces positive predictions of $q'$ and incorporates a physics-informed monotonicity constraint directly into the model. Symmetry considerations imply that $q'(\rho=0)=0$ at the magnetic axis, which can be omitted so that model predicts 100 $q'$ values instead of 101 $q$ values. Predicting $q'$ instead of $q$ is therefore advantageous because it directly incorporates known profile behavior while also representing a related quantity magnetic shear.

\begin{figure}[!htb]
\begin{center}
\includegraphics[width=\columnwidth]{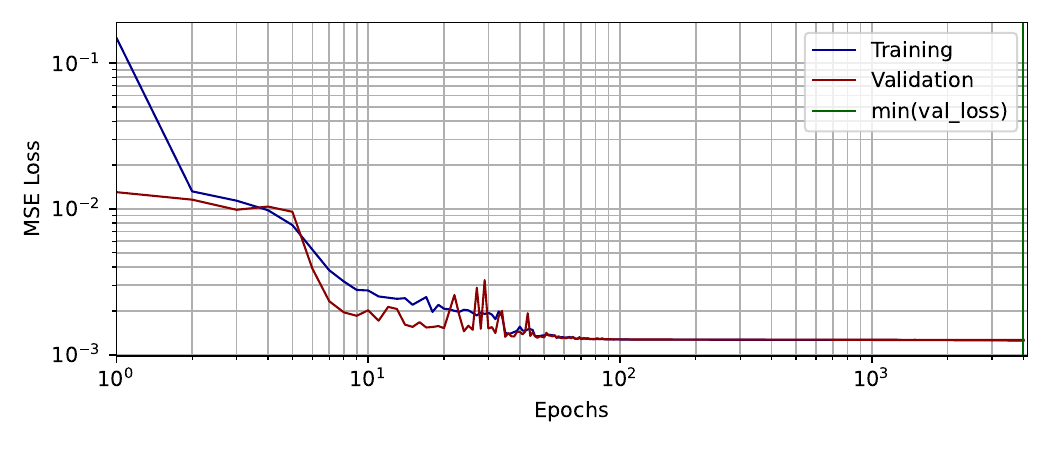}
\end{center}
\caption{MSE Loss for Training and Validation Datasets for model shown in Table \ref{tbl:q_cnn_model}.}
\label{fig:q_cnn_loss}
\end{figure}

The target $q'$ profiles were computed from the original $q$ profiles using the central-difference formula, except at the $\rho=1$ where backward-difference was used. The complete $q$ profile can be reconstructed through backward integration using $q_1$ predicted by the model described in Section \ref{sec:q1}. Because the profile is reconstructed through cumulative integration from the plasma edge toward the magnetic axis, prediction errors may accumulate and therefore have a larger impact on core values than on edge values. To verify the monotonicity assumption, all 100,760 equilibrium cases in the dataset were examined. Only 37 cases had slightly negative values of $q'$ at a few points near the magnetic axis, likely arising from numerical errors. These cases were excluded from the training dataset for the present model. After the \texttt{keras-tuner} hyper-optimization, the best-performing model found is shown in Table \ref{tbl:q_cnn_model}. The plot of MSE loss for training and validation datasets with respect to epochs is shown in Figure \ref{fig:q_cnn_loss}.

{\centering
\begin{longtblr}{
    colspec  = {l l l},
    rowhead  = 0,        
    rowfoot  = 0,        
  }
\toprule
Layer & Specification & \small{Activation} \\
\midrule
Input    & 28 Units & --- \\
\midrule
Dense & 160 Units & softplus \\
Dense & 192 Units & silu \\
Dense & 3200 Units & relu \\
Reshape & Shape $3200\rightarrow 100\times32$ & --- \\
Conv1D & \footnotesize{Filters 16, Kernel Size 9, Strides 1} & relu \\
Conv1D & \footnotesize{Filters 48, Kernel Size 3, Strides 1} & relu \\
Conv1D & \footnotesize{Filters 1, Kernel Size 3, Strides 1} & linear \\
Flatten & Shape $\rightarrow100$ & --- \\
\midrule
Output & 100 Units & softplus \\
\bottomrule
\end{longtblr}
\addtocounter{table}{-1}
\captionof{table}{1d-CNN Model to predict $q'$ profile. Total trainable parameters: 670,373.}
\label{tbl:q_cnn_model}
}

Similar to Section \ref{sec:q_pca}, the predictive performance of the model on the test dataset is summarized in Table \ref{tbl:q_cnn_performance}, which reports the $95^{th}$ and $99^{th}$ percentiles of the absolute prediction error for $q'$ and $q$ values at selected $\rho$ points. Comparison with the PCA-based model discussed in Section \ref{sec:q_pca} indicates that the 1d-CNN approach has similar accuracy near the core region but achieves slightly improved accuracy towards the plasma edge. However, a direct comparison should be interpreted cautiously because reconstructing $q$-profile in the CNN approach explicitly incorporates the separately predicted $q_1$, whereas the PCA model predicts the complete profile without it.

\begin{figure}[!htb]
\begin{center}
\includegraphics[width=\columnwidth]{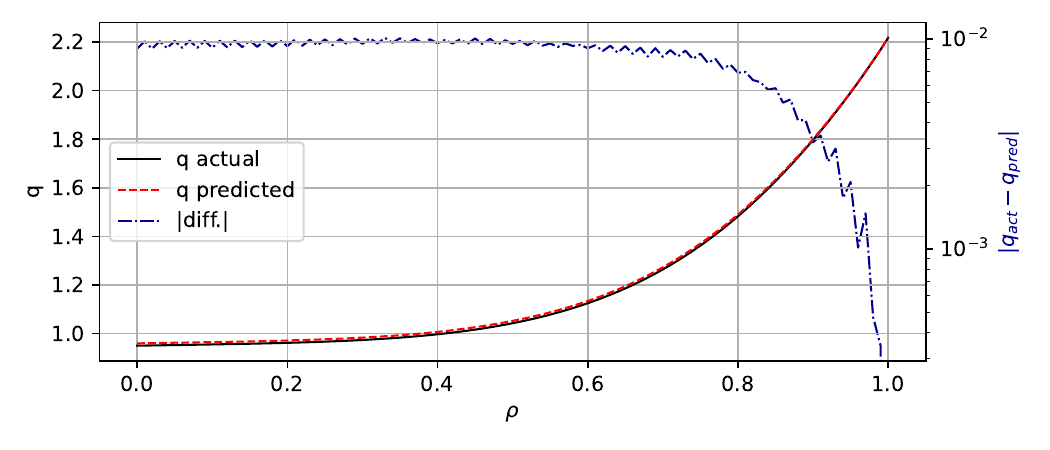}
\end{center}
\caption{Predicted $q$-profile with median MSE, for model shown in Table \ref{tbl:q_cnn_model}. Values on left axis and absolute difference on right axis.}
\label{fig:q_cnn_median}
\end{figure}

\begin{figure}[!htb]
\begin{center}
\includegraphics[width=\columnwidth]{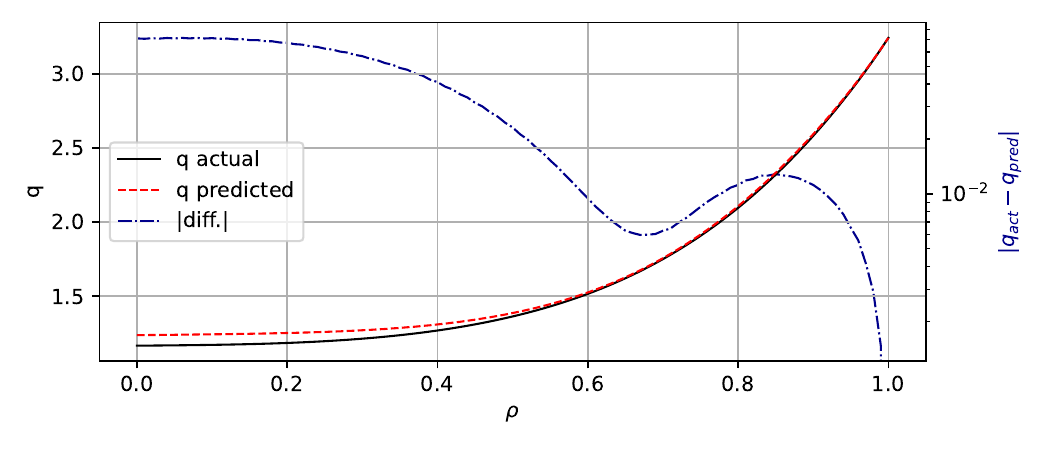}
\end{center}
\caption{Predicted $q$-profile with $99^{th}$ percentile in MSE, for model shown in Table \ref{tbl:q_cnn_model}. Values on left axis and absolute difference on right axis.}
\label{fig:q_cnn_99p}
\end{figure}

\begin{figure}[!htb]
\begin{center}
\includegraphics[width=\columnwidth]{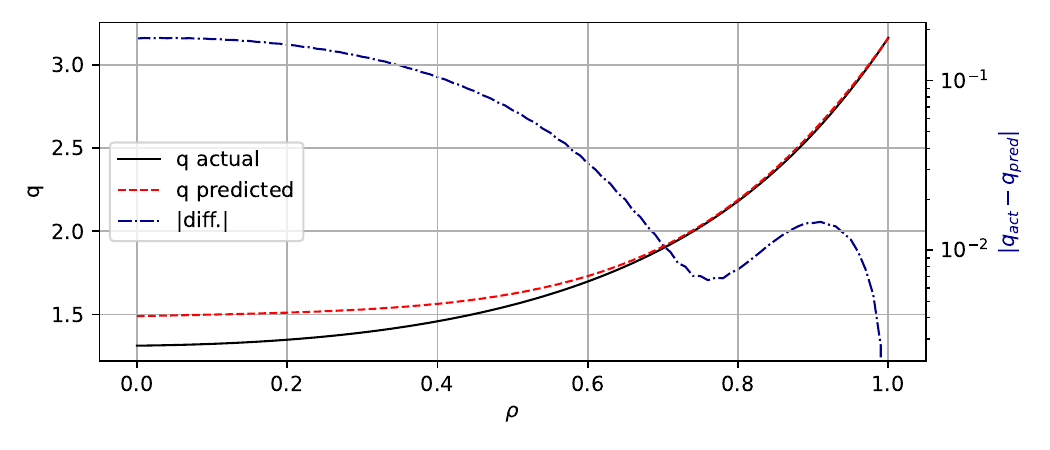}
\end{center}
\caption{Predicted $q$-profile with highest MSE, for model shown in Table \ref{tbl:q_cnn_model}. Values on left axis and absolute difference on right axis.}
\label{fig:q_cnn_worst}
\end{figure}

When test dataset predictions are sorted by MSE, the median, $99^{th}$ percentile and worst predictions are shown in Figures \ref{fig:q_cnn_median}, \ref{fig:q_cnn_99p} and \ref{fig:q_cnn_worst}, respectively. Even for the worst prediction in the test dataset, the error in the core safety factor is approximately 13\%, while the edge region remains well reproduced. The largest discrepancies generally occur near the magnetic axis, where the $q$ is particularly sensitive to subtle variations in the $J_\phi$ profile. Thus small uncertainties in the inferred $J_\phi$ profile by the model can therefore produce comparatively larger errors near core region than the edge. Overall, the results demonstrate that the combination of CNN-based local feature learning and physics-informed monotonicity constraints provides an effective framework for modeling ADITYA-U $q$-profiles.

{\centering
\begin{longtblr}{
    colspec  = {p{0.15\linewidth} p{0.12\linewidth} p{0.26\linewidth} p{0.26\linewidth}},
    rowhead  = 0,        
    rowfoot  = 0,        
  }
\toprule
Variable & Data Range & \footnotesize{$95^{th}$ percentile in Absolute Error} & \footnotesize{$99^{th}$ percentile in Absolute Error} \\
\midrule
$q'$($\rho$=0.01) & \footnotesize{2.64e-3, 0.132} & 9.58e-3 & 1.36e-2 \\
$q'$($\rho$=0.1) & \footnotesize{4.30e-3, 0.180} & 6.71e-3 & 1.34e-2 \\
$q'$($\rho$=0.9) & \small{1.95, 7.54} & 3.63e-2 & 5.74e-2 \\
$q'$($\rho$=1) & \small{3.16, 11.0} & 8.64e-2 & 0.139 \\
\midrule
$q$($\rho$=0) & \small{0.900, 1.50} & 4.33e-2 & 6.97e-2 \\
$q$($\rho$=0.1) & \small{0.901, 1.51} & 4.38e-2 & 7.01e-2 \\
$q$($\rho$=0.9) & \small{1.64, 3.00} & 6.19e-3 & 1.00e-2 \\
$q$($\rho$=0.99) & \small{1.96, 3.72} & 8.64e-4 & 1.39e-3 \\
\bottomrule
\end{longtblr}
\addtocounter{table}{-1}
\captionof{table}{Performance in predictions of $q'$ and $q$ at some $\rho$ values, for test dataset using the model shown in Table \ref{tbl:q_cnn_model}.}
\label{tbl:q_cnn_performance}
}

\section{Poloidal Flux Profile Models}
\label{sec:psi}
The equilibrium database contains poloidal flux $\psi(R,Z)$ profiles on a $71\times71$ rectangular grid. To predict the 2d arrays of $\psi$, PCA and 2d-CNN methods were used. The $\psi$ profiles in the generated dataset exhibit a relatively restricted family of shapes approximately characterized by nested concentric circular flux surfaces. This suggests that the dominant variation in the dataset may be represented efficiently using a low-dimensional basis obtained through PCA method. The strong local spatial correlations make 2d-CNN a natural choice for direct profile reconstruction. Similar to Section \ref{sec:q}, performing both strategies on the same dataset also allows to make a direct comparison between their performances. The PCA-based approach is described in Section \ref{sec:psi_pca} and the 2d-CNN approach is presented in Section \ref{sec:psi_cnn}. In addition to the input variables used in Section \ref{sec:q}, the quantities $q_1$, $\psi_{axs}$, and $\psi_{lim}$ were also included as model inputs, which in practical deployment can be used from the results of Sections \ref{sec:q1} and \ref{sec:psi01}.

Since $\psi$ is governed by the Grad-Shafranov equation, a purely data-driven model may produce flux distributions that fit the training data but do not remain fully consistent with the underlying equilibrium physics. To encourage physically consistent predictions, both the PCA and CNN models were formulated as PINNs by constraining the training with equation \ref{eq:gs}. In addition to predicting $\psi$, the models also predict the $J_\phi$ profile parameters $\alpha$, $\beta$, and $\gamma$, which are used to reconstruct the $J_{\phi}$ profile during the training, allowing the computation of mean-squared Grad-Shafranov residual $\delta GS=\langle(\Delta^*\psi+\mu_0RJ_\phi)^2\rangle$. The quantity $\delta GS$ is incorporated as an additional term in the training loss function, thereby penalizing solutions that deviate from the Grad-Shafranov equation. Although minimizing $\delta GS$ alone does not guarantee a physically valid equilibrium solution, it acts as a physics-based regularization mechanism and discourages predictions that are inconsistent with the governing equilibrium equation.

\subsection{PCA Model for $\psi$-profile}
\label{sec:psi_pca}
\begin{figure}[!htb]
\begin{center}
\includegraphics[width=\columnwidth]{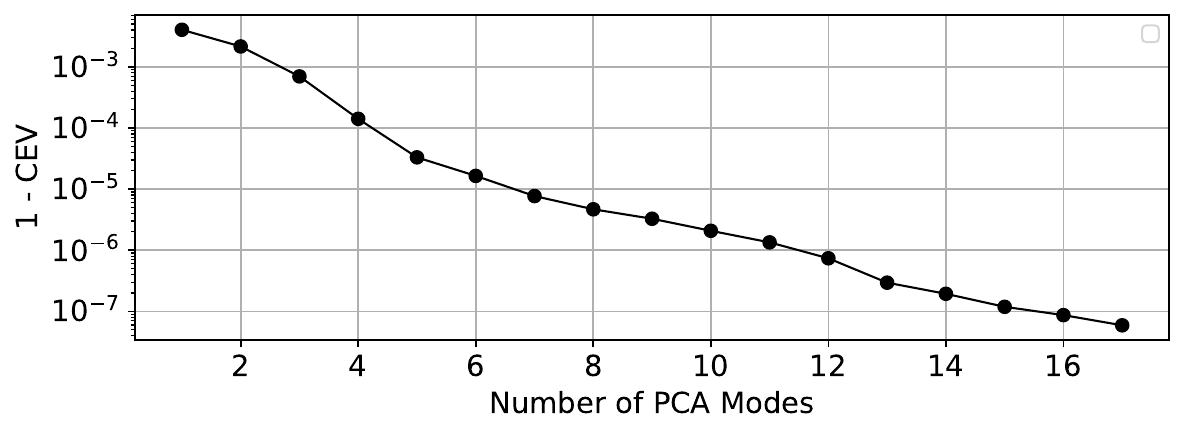}
\end{center}
\caption{$1-CEV$ with respect to number of retained PCA modes for the $\psi$-profiles.}
\label{fig:psi_pca_cev}
\end{figure}

\begin{figure}[!htb]
\begin{center}
\includegraphics[width=\columnwidth]{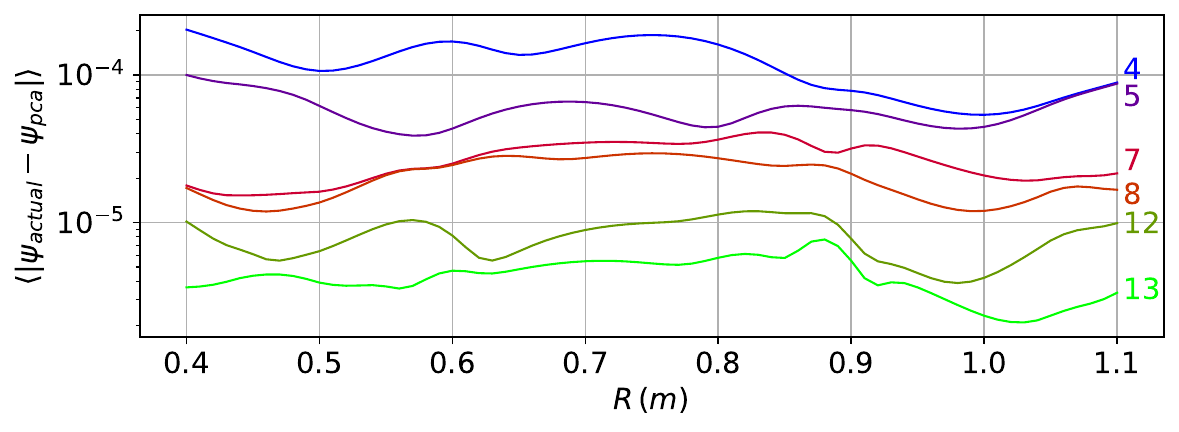}
\end{center}
\caption{Mean absolute difference between actual and reconstructed $\psi$-profile for different number of retained PCA modes. Mean has been taken over all cases and across $Z$-axis. The colored numbers on right side denote the numbers of PCA modes.}
\label{fig:psi_pca_err}
\end{figure}

Similar to the analysis performed for the $q$-profiles in Section \ref{sec:q_pca}, the cumulative explained variance (CEV) was evaluated for different numbers of PCA modes applied to the $\psi$ profiles in the dataset. Figure \ref{fig:psi_pca_cev} shows the variation of $1-CEV$ with the number of retained PCA modes and Figure \ref{fig:psi_pca_err} shows the mean absolute reconstruction error $\langle|\psi_{actual}-\psi_{pca}|\rangle$, averaged over all cases and the $Z$ direction. The results indicate that a CEV of 99.99\% is achieved using only 5 PCA modes, demonstrating that the dominant variability of $\psi$ profiles can be represented in a highly compressed form. However, unlike the safety-factor profile reconstruction problem, accurate estimation of $\psi$ is also important for evaluating $\Delta^*\psi$. Thus retaining only the minimum number of modes required to achieve high CEV may not be sufficient to accurately reproduce finer spatial variations in the $\psi$ profiles. For this reason, additional PCA models containing 8 and 13 modes were also investigated. The 13-mode representation was selected because Figures \ref{fig:psi_pca_cev} and \ref{fig:psi_pca_err} show a noticeable improvement in both CEV and $\langle|\psi_{actual}-\psi_{pca}|\rangle$ compared to the preceding 12-mode representation.

\begin{figure}[!htb]
\begin{center}
\includegraphics[width=\columnwidth]{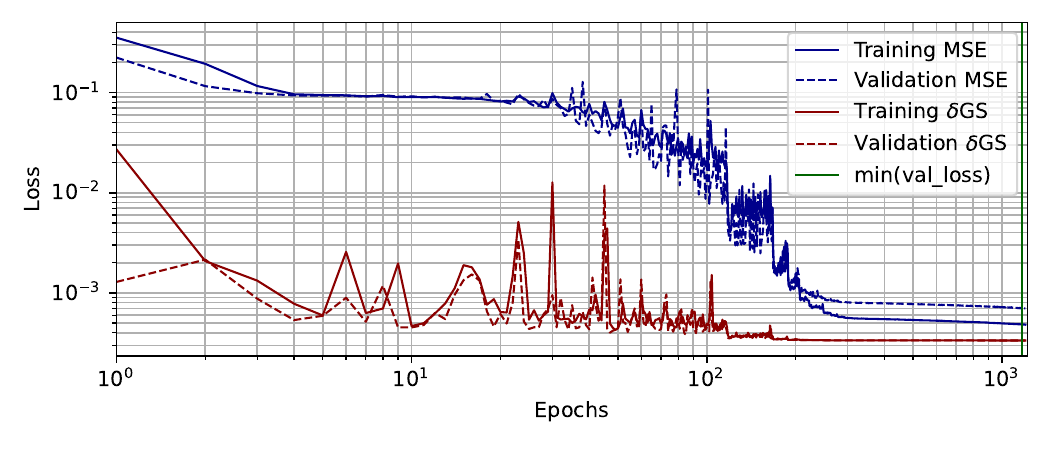}
\end{center}
\caption{MSE and $\delta GS$ Losses for Training and Validation Datasets for model shown in Table \ref{tbl:psi_pca_model}.}
\label{fig:psi_pca_loss}
\end{figure}

The PCA coefficients and the profile parameters $\alpha,\beta,\gamma$ were modeled simultaneously using a dense neural network. Based on the typical magnitudes of the MSE and Grad-Shafranov residual, the total training loss was chosen as $MSE+2\cdot\delta GS$. Models containing 5, 8, and 13 PCA modes were trained and evaluated. Unlike the behavior observed for higher-order PCA modes of the $q$ profiles in Section \ref{sec:q_pca}, the higher-order $\psi$ PCA modes remained sufficiently predictable. Among these, the 13-mode model consistently provided the best overall performance and is therefore presented in the remainder of this section. The best-performing model after doing the \texttt{keras-tuner} hyper-optimization is shown in Table \ref{tbl:psi_pca_model}. The plot of MSE and $\delta GS$ losses for training and validation datasets with respect to epochs is shown in Figure \ref{fig:psi_pca_loss}. The rapid loss in $\delta GS$ during the initial training stages indicates that model quickly learns to reconstruct $\psi$ profile consistent with the Grad-Shafranov equation. Although the validation MSE remains moderately higher than the training MSE during later epochs, the gap remains bounded and no significant degradation in test dataset prediction quality is observed.

{\centering
\begin{longtblr}{
    colspec  = {l l l},
    rowhead  = 0,        
    rowfoot  = 0,        
  }
\toprule
Layer & Specification & Activation \\
\midrule
Input    & 31 Units & --- \\
\midrule
Dense & 512 Units & gelu \\
Dense & 1536 Units & gelu \\
Dense & 1152 Units & gelu \\
Dense & 1280 Units & relu \\
\midrule
Output & 16 Units & linear \\
\bottomrule
\end{longtblr}
\addtocounter{table}{-1}
\captionof{table}{Dense Model to predict 13 PCA modes of $\psi$-profile and parameters $\alpha,\beta,\gamma$. Total trainable parameters: 4,071,312.}
\label{tbl:psi_pca_model}
}

\begin{figure}[!htb]
\begin{center}
\includegraphics[width=\columnwidth]{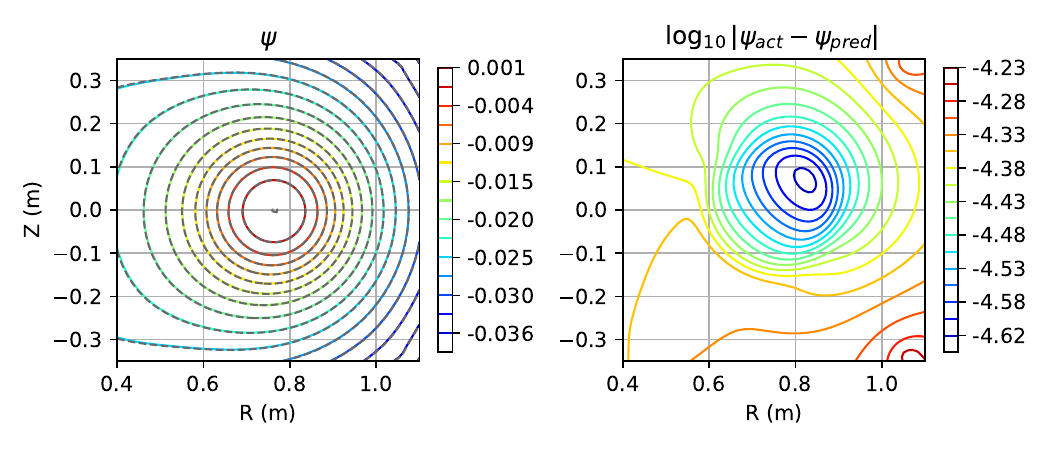}
\end{center}
\caption{Predicted $\psi$-profile with median MSE, for model shown in Table \ref{tbl:psi_pca_model}. In left plot, colored contours show the actual values and dashed gray contours show predicted values. Right plot shows $\log_{10}$ of absolute difference between actual and predicted values.}
\label{fig:psi_pca_median}
\end{figure}

\begin{figure}[!htb]
\begin{center}
\includegraphics[width=\columnwidth]{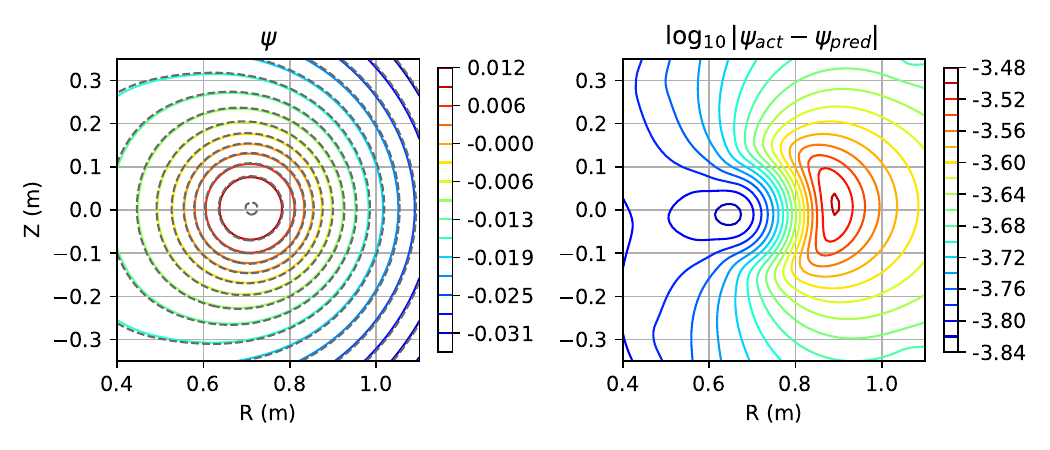}
\end{center}
\caption{Predicted $\psi$-profile with $99^{th}$ percentile in MSE, for model shown in Table \ref{tbl:psi_pca_model}. In left plot, colored contours show the actual values and dashed gray contours show predicted values. Right plot shows $\log_{10}$ of absolute difference between actual and predicted values.}
\label{fig:psi_pca_99p}
\end{figure}

\begin{figure}[!htb]
\begin{center}
\includegraphics[width=\columnwidth]{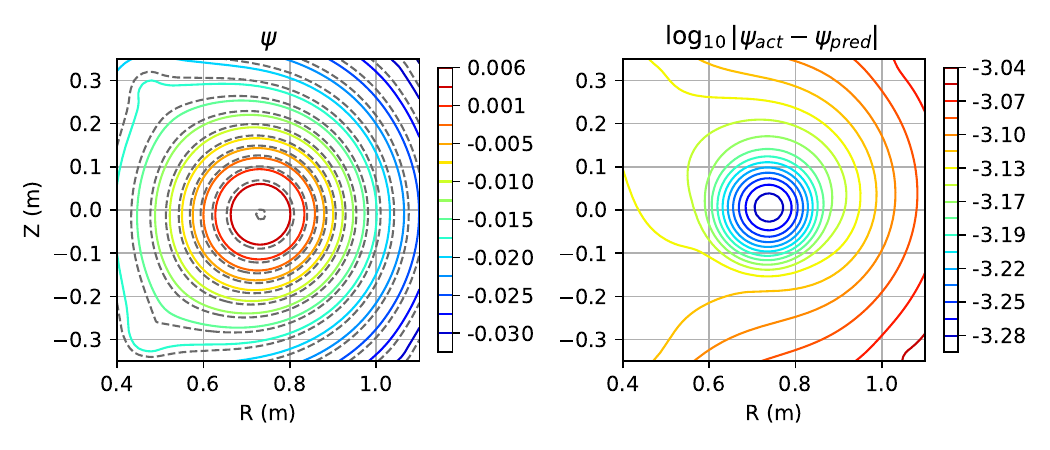}
\end{center}
\caption{Predicted $\psi$-profile with highest MSE, for model shown in Table \ref{tbl:psi_pca_model}. In left plot, colored contours show the actual values and dashed gray contours show predicted values. Right plot shows $\log_{10}$ of absolute difference between actual and predicted values.}
\label{fig:psi_pca_worst}
\end{figure}

The predictive performance of the model on the test dataset is summarized in Table \ref{tbl:psi_pca_performance}, which reports the $95^{th}$ and $99^{th}$ percentiles in absolute difference between actual and predicted values of 13 PCA modes and $\alpha,\beta,\gamma$ values. When test dataset predictions are sorted by MSE, the median, $99^{th}$ percentile, and worst-case predictions are shown in Figures \ref{fig:psi_pca_median}, \ref{fig:psi_pca_99p} and \ref{fig:psi_pca_worst}, respectively. The predicted flux distributions remain smooth and physically consistent across these representative cases, indicating that the PCA-based PINN framework is capable of accurately reconstructing the dominant equilibrium structure of the $\psi$ profiles.

{\centering
\begin{longtblr}{
    colspec  = {p{0.16\linewidth} p{0.13\linewidth} p{0.26\linewidth} p{0.26\linewidth}},
    rowhead  = 0,        
    rowfoot  = 0,        
  }
\toprule
Variable & Data Range & \footnotesize{$95^{th}$ percentile in Absolute Error} & \footnotesize{$99^{th}$ percentile in Absolute Error} \\
\midrule
PCA Mode 1 & \small{-3.4865, 2.6851} & 8.138e-3 & 1.487e-2 \\
PCA Mode 2 & \small{-0.1445, 0.2113} & 7.993e-4 & 1.658e-3 \\
PCA Mode 3 & \small{-0.1524, 0.1835} & 7.748e-4 & 2.145e-3 \\
PCA Mode 4 & \small{-0.0938, 0.0558} & 3.385e-4 & 6.802e-4 \\
PCA Mode 5 & \small{-0.0369, 0.0383} & 1.768e-4 & 3.671e-4 \\
PCA Mode 6 & \small{-0.0264, 0.0184} & 7.530e-5 & 1.450e-4 \\
PCA Mode 7 & \small{-0.0066, 0.0314} & 4.364e-5 & 9.034e-5 \\
PCA Mode 8 & \small{-0.0176, 0.0044} & 4.004e-5 & 7.540e-5 \\
PCA Mode 9 & \small{-0.0204, 0.0102} & 2.334e-5 & 4.410e-5 \\
PCA Mode 10 & \small{-0.0103, 0.0098} & 2.524e-5 & 4.909e-5 \\
PCA Mode 11 & \small{-0.0037, 0.0064} & 3.540e-5 & 6.469e-5 \\
PCA Mode 12 & \small{-0.0076, 0.0024} & 3.581e-5 & 6.843e-5 \\
PCA Mode 13 & \small{-0.0040, 0.0062} & 1.708e-5 & 3.368e-5 \\
\midrule
$\alpha$ & \small{2.2506, 5.6443} & 4.355e-2 & 8.126e-2 \\
$\beta$ & \small{0.0618, 0.4430} & 8.692e-4 & 1.497e-3 \\
$\gamma$ & \small{1.5122, 8.9000} & 1.411e-1 & 2.644e-1 \\
\bottomrule
\end{longtblr}
\addtocounter{table}{-1}
\captionof{table}{Performance in predictions of 13 PCA modes and $\alpha,\beta,\gamma$ values, for test dataset using the model shown in Table \ref{tbl:psi_pca_model}.}
\label{tbl:psi_pca_performance}
}

\subsection{2d-CNN Model for $\psi$-profile}
\label{sec:psi_cnn}
In this approach, $\psi$-profiles are modeled directly using a 2d-CNN, where the $\psi$ profile is treated as an image of size $71\times71$. CNN architectures designed for image generation commonly construct an image through a sequence of progressive upsampling operations starting from a low-dimensional latent representation. Since 71 is a prime number, direct application of a conventional power-of-two upsampling strategy is not possible. Consequently, the network first generates a $3\times3$ feature map, which is upsampled with a transposed convolution layer to $9\times9$ and subsequently with three more upsampling stages to produce a $72\times72$ image. The final output is cropped to obtain the required $\psi$ profile of size $71\times71$.

\begin{figure}[!htb]
\begin{center}
\includegraphics[width=\columnwidth]{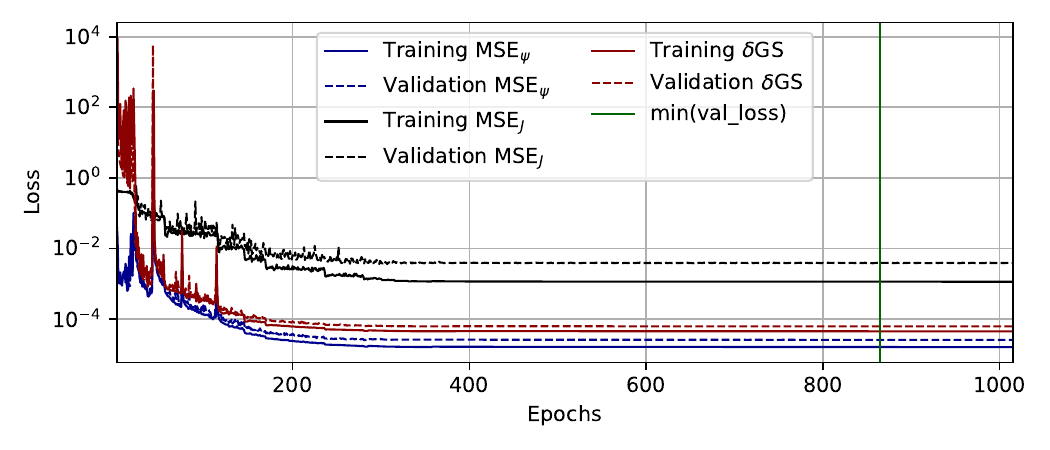}
\end{center}
\caption{$\text{MSE}_\psi$, $\text{MSE}_J$ and $\delta GS$ Losses for Training and Validation Datasets for model shown in Table \ref{tbl:psi_cnn_model}.}
\label{fig:psi_cnn_loss}
\end{figure}

The model simultaneously predicts $\psi$ and $\alpha,\beta,\gamma$ parameters. Thus this model has three different kinds of loss functions: $\text{MSE}_\psi=$ MSE of 5041 $\psi$ values, $\text{MSE}_J=$ MSE of $\alpha,\beta,\gamma$ parameters and $\delta GS=$ Grad-Shafranov residual. Direct incorporation of these losses into a single loss function was found to produce unstable training behavior. In particular, the initial values of $\delta GS$ for randomly initialized networks were typically several orders of magnitude larger than the other two MSE. In an attempt to only lower $\delta GS$, the training quickly converges to model weights that represent inappropriate solutions despite having low $\delta GS$. By the time training starts to consider $\text{MSE}_\psi$ and $\text{MSE}_J$, the model weights are trapped in rather distorted solutions that training struggles to further improve on these MSE, resulting in a higher overall loss despite low $\delta GS$.To mitigate this issue, the contribution of the Grad-Shafranov residual was activated progressively during training. A multiplicative factor $f$ was introduced such that the influence of $\delta GS$ remained negligible when $\text{MSE}_\psi$ and $\text{MSE}_J$ were large and gradually increased as the reconstruction quality improved. A shifted sharp inverse sigmoid function, $f=sigmoid(-50(\text{MSE}_\psi + \text{MSE}_J - 0.15))$ was found to provide the satisfactory behavior. This formulation allows the network to first learn the overall structure of the $\psi$ profiles before $\delta GS$ becomes a significant component of the total loss function. An additional challenge arises from the simultaneous use of three loss terms with substantially different magnitudes and optimization characteristics. Rather than manually selecting weighting coefficients, dynamic loss balancing based on the uncertainty-weighting approach of \cite{2018Kendall} was adopted. This method introduces trainable parameters $\sigma_{\psi}$, $\sigma_{J}$, and $\sigma_{GS}$ that automatically adjust the relative contributions of the three loss components during the training. The resulting total loss function is given by
\begin{equation}
e^{-\sigma_\psi}\text{MSE}_\psi + \sigma_\psi + e^{-\sigma_J}\text{MSE}_J + \sigma_J + f\left(e^{-\sigma_{GS}}\delta GS + \sigma_{GS}\right)
\end{equation}
After the \texttt{keras-tuner} hyper-optimization, the best-performing model found is shown in Table \ref{tbl:psi_cnn_model}. The plot of MSE losses and $\delta GS$ for training and validation datasets with respect to epochs is shown in Figure \ref{fig:psi_cnn_loss}.

{\centering
\footnotesize
\begin{longtblr}{
    colspec  = {p{0.24\linewidth} p{0.45\linewidth} p{0.15\linewidth}},
    rowhead  = 0,        
    rowfoot  = 0,        
  }
\toprule
Layer & Specification & \small{Activation} \\
\midrule
Input    & 31 Units & --- \\
\midrule
Dense & 512 Units & silu \\
Dense & 1024 Units & silu \\
Dense & 1024 Units & gelu \\
Dense & 256 Units & relu \\
\midrule
\SetCell[c=3]{c} \textit{Branch 1: 2d-CNN for $\psi$} \\
\midrule
Dense & 864 Units & relu \\
Reshape & Shape $864\rightarrow 3\times3\times96$ & --- \\
Conv2DTranspose & Filters 64, Kernel Size (7,7), Strides (3,3) & elu \\
Conv2DTranspose & Filters 160, Kernel Size (5,5), Strides (2,2) & elu \\
Conv2DTranspose & Filters 80, Kernel Size (5,5), Strides (2,2) & elu \\
Conv2DTranspose & Filters 16, Kernel Size (3,3), Strides (2,2) & elu \\
Cropping2D & $(0,1)$, $(0,1)$ & --- \\
Conv2D & Filters 1, Kernel Size (3,3), Strides (1,1) & linear \\
\midrule
\SetCell[c=3]{c} \textit{Branch 2: Dense layers for $\alpha,\beta,\gamma$} \\
\midrule
Dense & 512 Units & silu \\
Dense & 1024 Units & silu \\
Dense & 1024 Units & gelu \\
Dense & 3 Units & linear \\
\bottomrule
\end{longtblr}
\addtocounter{table}{-1}
\captionof{table}{2d-CNN Model to predict $\psi$ profile. Total trainable parameters: 4,674,356.}
\label{tbl:psi_cnn_model}
}

\begin{figure}[!htb]
\begin{center}
\includegraphics[width=\columnwidth]{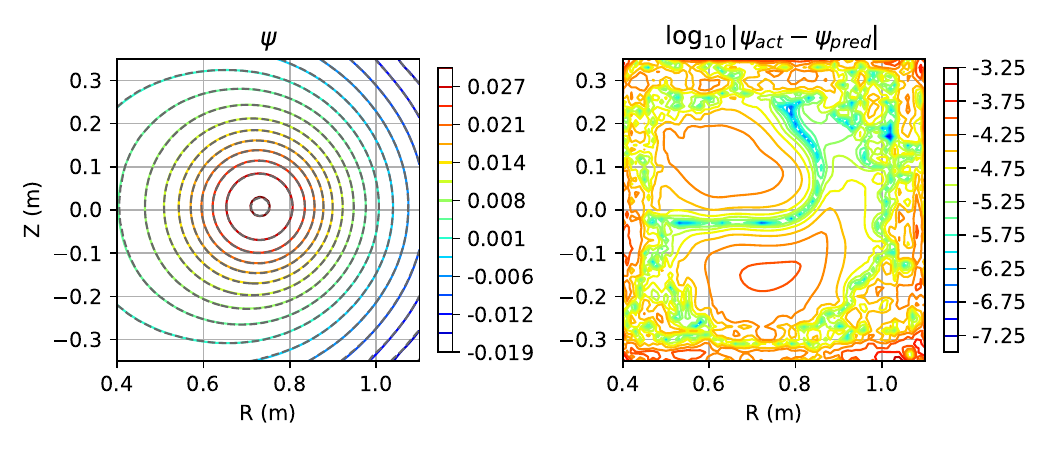}
\end{center}
\caption{Predicted $\psi$-profile with median MSE, for model shown in Table \ref{tbl:psi_cnn_model}. In left plot, colored contours show the actual values and dashed gray contours show predicted values. Right plot shows $\log_{10}$ of absolute difference between actual and predicted values.}
\label{fig:psi_cnn_median}
\end{figure}

\begin{figure}[!htb]
\begin{center}
\includegraphics[width=\columnwidth]{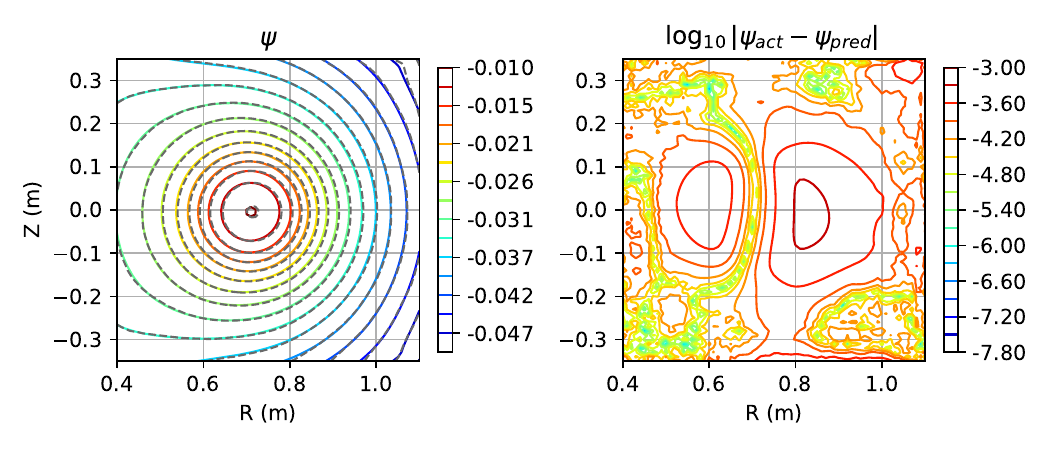}
\end{center}
\caption{Predicted $\psi$-profile with $99^{th}$ percentile in MSE, for model shown in Table \ref{tbl:psi_cnn_model}. In left plot, colored contours show the actual values and dashed gray contours show predicted values. Right plot shows $\log_{10}$ of absolute difference between actual and predicted values.}
\label{fig:psi_cnn_99p}
\end{figure}

\begin{figure}[!htb]
\begin{center}
\includegraphics[width=\columnwidth]{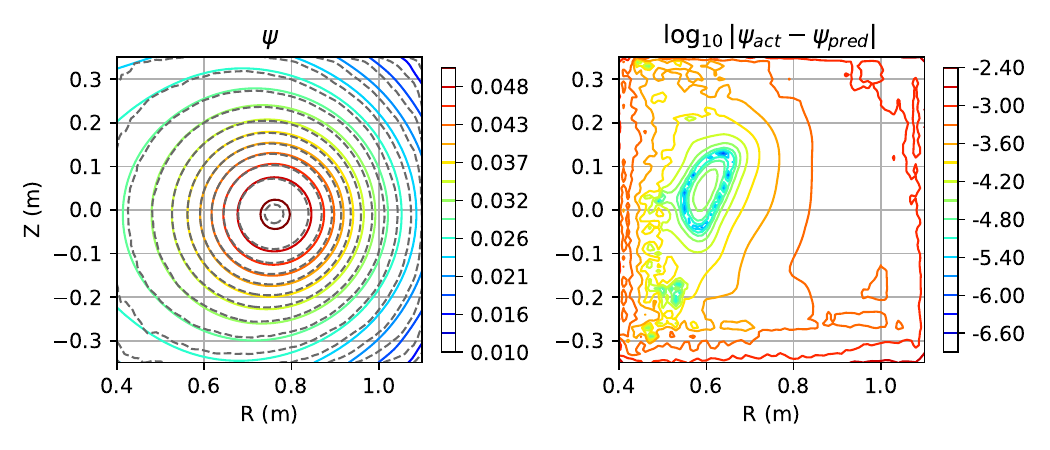}
\end{center}
\caption{Predicted $\psi$-profile with highest MSE, for model shown in Table \ref{tbl:psi_cnn_model}. In left plot, colored contours show the actual values and dashed gray contours show predicted values. Right plot shows $\log_{10}$ of absolute difference between actual and predicted values.}
\label{fig:psi_cnn_worst}
\end{figure}

The predictive performance of the model on the test dataset is summarized in Table \ref{tbl:psi_cnn_performance}, which reports the $95^{th}$ and $99^{th}$ percentiles in absolute prediction errors for selected $\psi$ values and for $\alpha,\beta,\gamma$ parameters. When test dataset predictions are sorted by MSE, the median, $99^{th}$ percentile, and worst-case predictions are shown in Figures \ref{fig:psi_cnn_median}, \ref{fig:psi_cnn_99p} and \ref{fig:psi_cnn_worst}, respectively. Comparison with the PCA-based results presented in Figures \ref{fig:psi_pca_median}-\ref{fig:psi_pca_worst} reveals notable differences in the error characteristics of the two approaches. The CNN model generally achieves superior overall reconstruction accuracy, however the resulting errors may be spatially localized, producing localized distortions in portions of the flux distribution. In contrast, the PCA model tends to distribute reconstruction errors more globally across the profile, resulting in smoother reconstructions that more closely preserve the dominant structure of $\psi$ profiles. 

{\centering
\begin{longtblr}{
    colspec  = {p{0.16\linewidth} p{0.12\linewidth} p{0.26\linewidth} p{0.26\linewidth}},
    rowhead  = 0,        
    rowfoot  = 0,        
  }
\toprule
Variable & Data Range & \footnotesize{$95^{th}$ percentile in Absolute Error} & \footnotesize{$99^{th}$ percentile in Absolute Error} \\
\midrule
$\psi(0.5,0)$ & \footnotesize{-5.4e-2, 3.5e-2} & 1.039e-04 & 1.763e-04 \\
$\psi(0.75,0)$ & \footnotesize{-3.7e-2, 5.2e-2} & 6.651e-05 & 1.269e-04 \\
$\psi(1.0,0)$ & \footnotesize{-6.2e-2, 3.2e-2} & 1.079e-04 & 1.814e-04 \\
\footnotesize{$\psi(0.41,0.34)$} & \footnotesize{-6.1e-2, 2.4e-2} & 2.158e-04 & 3.160e-04 \\
\footnotesize{$\psi(0.75,0.34)$} & \footnotesize{-6.4e-2, 2.6e-2} & 1.087e-04 & 1.944e-04 \\
\footnotesize{$\psi(1.09,0.34)$} & \footnotesize{-7.6e-2, 1.5e-2} & 2.259e-04 & 3.243e-04 \\
\midrule
$\alpha$ & \small{2.3633, 5.8475} & 4.000e-02 & 6.904e-02 \\
$\beta$ & \small{0.0586, 0.4922} & 6.529e-04 & 1.071e-03 \\
$\gamma$ & \small{1.4484, 8.8986} & 0.1268 & 0.2194 \\
\bottomrule
\end{longtblr}
\addtocounter{table}{-1}
\captionof{table}{Performance in predictions of $\psi$ values at some points and $\alpha,\beta,\gamma$ values, for test dataset using the model shown in Table \ref{tbl:psi_cnn_model}.}
\label{tbl:psi_cnn_performance}
}

\section{Conclusion}
\label{sec:end}
This work presents a comprehensive study of deep learning models for ADITYA-U MHD equilibrium using a large synthetic free-boundary equilibrium database generated with the pyIPREQ code. The dataset was derived from experimentally motivated operating conditions obtained from 766 ADITYA-U discharges and constrained using physics-informed filtering criteria to retain realistic circular limiter equilibria close to the flat-top phase, as described in Section \ref{sec:dtgen}. Multiple deep learning strategies were investigated for predicting scalar equilibrium quantities, 1d safety factor profiles and 2d poloidal flux profiles. The study explored dense neural networks, PCA-based reduced-order models, 1d and 2d CNN models, and physics-informed formulations incorporating Grad-Shafranov residual. The results demonstrate that the constrained operational space considered in this work admits effective reduced-order representations and enables accurate surrogate modeling of key equilibrium quantities using deep learning techniques.

Several important observations emerge from the present study. Scalar equilibrium quantities, including the magnetic-axis position, edge safety factor, poloidal beta, and normalized internal inductance, were predicted with good accuracy using compact Dense neural-network models. The inverse coil-current model demonstrated that operationally meaningful actuator trends can be inferred from desired plasma parameters within the operating domain represented by the dataset. PCA-based approaches revealed that both $q$ and $\psi$ profiles occupy relatively low-dimensional manifolds for the considered equilibrium regime, whereas CNN-based models were more effective at capturing local spatial features. The comparison between PCA and CNN approaches highlighted that PCA models generally produced smoother reconstructions that preserved the dominant equilibrium structure, whereas CNN models achieved improved local accuracy but were more susceptible to localized distortions. The incorporation of Grad-Shafranov residual losses provided an additional physics-based regularization mechanism for the $\psi$ models and helped discourage physically inconsistent predictions. The study also revealed that $J_\phi$ profile shape parameters $\alpha,\gamma$ and quantities more sensitive to them like $\ell_i$ and $q_0$ were more difficult to predict, suggesting their insufficient information in the available inputs. Further improvements may therefore require additional diagnostics that provide better information about $J_\phi$ profile. To assess computational efficiency of the models, the inference time of the largest model shown in Table \ref{tbl:psi_cnn_model} was evaluated on a multi-core CPU and found to be approximately \SI{1}{ms} per equilibrium instance, which should be further reducible using GPU and FPGA based implementations. This demonstrates that the developed models are computationally suitable for fast equilibrium analysis and have potential for real-time equilibrium estimation and plasma control applications in ADITYA-U operations.

The present work also has several limitations that motivate future developments. The generated equilibrium database is restricted to circular limiter plasmas and depends on the adopted current-density profile model, while a formal linear MHD stability analysis was not performed. The estimation of $\beta_p$ described in Section \ref{sec:betap} relied on a relatively simple probabilistic linear model derived from limited experimental data and could be improved through more advanced regression approaches or through the availability of more experimental data. Future studies may also investigate alternative reduced-order techniques, transformer-based architectures, uncertainty-aware models, and more advanced physics-informed formulations. Inclusion of additional diagnostic signals, eddy-current effects, and experimentally reconstructed equilibria would further improve the realism and applicability of the models. Since magnetic probe and loop voltage measurements were used as inputs to all forward models developed in this work, the resulting models are applicable only when all these measurements are reliably available. Consequently, these models should not be viewed as direct replacements of pyIPREQ solutions that are constructed without magnetic probe and loop voltage inputs, although the generated dataset provides a foundation for developing such models in future work. Nevertheless, to the authors' knowledge, this study constitutes the first integrated machine-learning framework for ADITYA-U MHD equilibrium modeling based on experimentally motivated operating scenarios. The methodology demonstrates the potential of combining physics-based equilibrium solvers with modern deep learning techniques to develop data-driven models for rapid equilibrium inference, plasma analysis, actuator planning, and real-time plasma control studies in tokamak operations.

\subsubsection*{Acknowledgment}
LLM AI Models have been used to rephrase the text of this article for better readability. This work was supported by the Institute for Plasma Research (IPR), India. Authors are thankful to ADITYA-U and data acquisition teams for conducting the experiments and making the data available. The computations for the work described in this article were performed on ANTYA, an IPR Linux Cluster.

\small
\bibliographystyle{unsrt}
\bibliography{main}

\end{document}